# Emerging spintronics phenomena and applications

Rahul Mishra and Hyunsoo Yang[*]

*Department of Electrical and Computer Engineering, National University of Singapore, 117576, Singapore*

Development of future sensor, memory, and computing nanodevices based on novel physical concepts is one of the significant research endeavors in solid-state research. The field of spintronics is one such promising area of nanoelectronics which utilizes both the charge and spin of an electron for device operations. The advantage offered by spin systems is in their non-volatility and low-power functionality. This paper reviews emerging spintronic phenomena and the research advancements in diverse spin based applications. Spin devices and systems for logic, memories, emerging computing schemes, flexible electronics and terahertz emitters are discussed in this report.

[*]eleyang@nus.edu.sg

# I. Introduction

Conventional sensor, memory, and computing electronics exploit the charge of an electron for their operations. However, along with charge, an electron is also characterized by its spin angular momentum or spin. It is the spin of an electron that manifests in the form of magnetism that we see in magnetic objects of the macro world. In the information technology age, magnetism has found industry applications in the massive digital data storage. The field of spintronics is centered on electron's spin in conjunction with its charge. As we near the end of a several decade scaling of CMOS technologies due to fundamental physical limitations, utilizing the degree of spin freedom might be a natural choice for next generation technologies. An external energy source is not required for maintaining a particular spin- or magnetic-state in a spintronic device. This property makes spintronic devices non-volatile and low power consuming, hence attractive for the emerging era of mobile, wireless and scaled down electronic applications.

The validation of spintronics as an imperative field to pursue was propelled with the discoveries of giant-magnetoresistance (GMR) [1,2] and tunneling magnetoresistance (TMR) [3,4]. In these findings it was put forward that a thin magnetic multilayer has a higher (lower) resistance when the magnetization of individual layers is aligned anti-parallel (parallel) to each other. Eventually, these structures were incorporated in the read head of hard disks, where their scalability has resulted in ultrahigh storage density. Presently, the read heads are based on the magnetic tunnel junction (MTJ) in which the spacer layer between the two ferromagnetic layers is typically an oxide such as MgO. While in a magnetic disk storage, the read-MTJ is toggled by the magnetic field emanating from the bits on the disk, a more efficient and scalable way to switch the magnetization in an MTJ-bit was found to be using the spin-transfer torque (STT) [5-10]. A STT involves a transfer of angular momentum from a moving electron to the magnetic atom. STT based devices have found recent commercial applications in magnetic random access memories (MRAM). Apart from their use in computing electronics, magnetic devices are also highly relevant in a wide variety of sensors.

Keeping in view the wide applications of the magnetics/spintronics till now, there has been an aggressive research effort in this field for improving the existing and development of the emerging applications. In this review, we will focus on the advancement of spintronics research in some of the major areas of technical and practical significance. In Section II, spintronic devices for logic computation will be discussed. Following this, in Section III, we will focus on spin devices for MRAM. In particular, the mechanism of spin-orbit torque (SOT) and subsequent research on the SOT devices will be detailed. Advancement in the spintronic devices and systems for alternative computing methodologies and optimization machines is part of Section IV. Progress on spin devices for flexible electronics and THz emitter are discussed in Section V and VI, respectively.

## II. Spintronics for logic

At present, most of the computational tasks are performed by the microprocessors. These computing units use miniaturized solid-state elements i.e. metal-oxide-semiconductor (MOS) to transfer and process the information. The information is transferred by the electronic charge and is stored in form of distinct voltage levels. The information processing capacity of a typical processor has been continuously increasing since its inception in early 1970's. This has been majorly due to aggressive scaling of the underlying complementary MOS (CMOS) technology which resulted in packing numerous functions in a given processor area [11]. However, the processor performance improvement has recently plateaued, partially due to increasing power density. Spintronic devices offer a potential solution to this problem by offering reduced power consumption owing to their non-volatile nature [12-14].

The Datta and Das transistor is among the very first proposed spin logic device [15,16]. The suggested structure as illustrated in Fig. 1(a) consists of a spin polarizer and an analyzer (a magnet) connected by a non-magnet which has spin-orbit coupling (SOC). The application of an electric field on the SOC channel results in precession of spin of the electrons that are injected from the polarizer. The precession of spin is a consequence of the Rashba effect which dictates the manifestation of electric field into a magnetic field in a moving electron's frame of reference [17]. The phase of the injected spin can therefore be controlled by a gate voltage applied on top of the channel. On arrival at the analyzer, if the electron's spin is in-phase (out-of-phase) with the analyzer's magnetization, it will result in a low (high) channel resistance corresponding to "On (Off)" state of the transistor. However, the use of phase makes it difficult for implementing logic circuits, because the phase is very sensitive with a continuous variable and not a discrete binary one. Therefore, phase information may be more suitable for sensors which deals with analog output values.

Till date a fully functional spin field-effect transistor (spin-FET) has not been realized due to non-ideality of the polarizer and analyzer layer, and the scattering process in solids which randomizes the spins [16]. However, there have been few demonstrations of gate induced spin precession and their subsequent detection in two-dimensional electron gas (2DEG) systems. For example, Koo et al. have detected spin precession in an InAs high-electron mobility channel [18]. NiFe electrodes were used as the spin injector and detector in this configuration (Fig. 1(b)). As shown in Fig. 1(c), an oscillatory modulation of channel conduction was observed when the magnetization of the polarizer and analyzer were aligned along the channel direction (black curve). It should be noted that with the magnetization aligned along an in-plane direction transverse to the channel there was no spin precession as the Rashba field and injected spins are collinear (red curve). It was later shown that spins injected using circularly polarized lights can also be modulated using a top gate and be detected using the spin Hall effect (SHE) [19]. Recently, spins injected electrically using a magnetic layer were detected in a similar way using the inverse spin Hall effect (ISHE) [20].

While the spin-FET emulates the functionality of a conventional MOSFET, numerous other spin logic devices which exploit alternate magnetic properties have been realized. One such scheme uses the magnetic domain wall (DW) and its motion to transfer information [21-24]. A magnetic DW is the interface between two oppositely aligned magnetic regions which moves either along or opposite to an applied magnetic field. The motion of DW around a corner in the presence of a rotating magnetic field has been used to implement NOT gate as shown in Fig. 2(a) [25]. A nanowire is patterned in the form of a cusp with its either side serving as input and output terminal of the NOT gate. The magnetization direction with respect to the DW motion direction represents the two logic states. A magnetic field is rotated in anti-clockwise direction as shown in Fig. 2(a). A DW that is present at a point P with the magnetization aligned along the +x direction on its left side (representing input of logic "1") moves along the first corner of the cusp to the point Q when the magnetic field is rotated from the +x to +y direction. In the next cycle of magnetic field rotation from the +y to –x direction the DW moves to the point R. However, on arrival at the point R, the direction of the magnetization of the left of DW is now towards –x direction due to the continuity of the magnetization in the cusp. A logic "1" input is thereby converted to logic "0" in a half-cycle of magnetic field rotation. The rotating magnetic field in effect serves as a clock signal. Input and output traces of the NOT gate obtained using the magneto-optic Kerr effect (MOKE) are shown in Fig. 2(b). It should be noted that there is a propagation delay of half clock cycle from input to output as explained previously. Shift registers were also implemented by pattering many such cusps adjacent to each other. Several other nano-patterning schemes have been used to implement NOT gate [23], AND gate [21], buffer [26], and SHIFT registers [24]. The DWs in these devices are either moved with the help of magnetic field or using spin current. A big disadvantage of the DW based logic is the need for external magnetic field to move the DW, which also prevent individual controllability of each DW in a device with multiple DWs. This hinders their scaling for practical applications. Ideally, the DW based logic gates can be designed such that the DWs can be moved by currents instead of magnetic fields. Another drawback of the DW based logic is the size of DW which can be anywhere between 7 to 100 nm. Therefore, the device containing these DW will be even larger. In comparison, the CMOS technology is already mass producing the 7-nm node technology and approaching the 5-nm node in the coming years. The DWs are also prone to pinning due to inhomogeneity of the patterned channel, which can cause reliability issues.

Another scheme of implementing spintronic logic is via switching of bistable magnetic element. The two stable states of a magnet represent binary information. Behin-Aein et al. proposed one such implementation in which spin information can be transferred using spin current from one magnetic element to another [27]. In their proposed scheme shown in Fig. 3(a), a voltage $V_{supply}$ is used to apply spin-torques to position the output magnetic bit in a neutral state (high-energy) which lies between the two stable states (low-energy). An application of a $V_{bias}$ signal which is relatively small compared to $V_{supply}$ transmits the information from the input magnet to the output magnet through the channel. A semiconductor spin channel can be used as an interconnect as it supports a longer spin coherence length. In the presence of both $V_{supply}$ and $V_{bias}$ the output magnet switches in either of the stable states depending on the state of the input bit. Information from few of these input magnets can be combined to implement logic functions like AND/OR. Figure 3(b) shows two cascading gates with two variable inputs and a fixed middle input. If the middle input is aligned along the logic "1" direction, the gate functions as an OR gate as the net spin current which is determined by the superposition of the spin currents from the 3 inputs will be in logic "1" direction if either of the variable input is "1". On the other hand, if the middle input is fixed in the "0" direction, the gate emulates AND functionality as the net spin current will be along logic "1" direction only when both the variable inputs are "1". The output terminal of the gate receives information when $V_{supply}$ is applied on it. The information is transferred to the next gate on an application of $V_{supply}$ in the next clock cycle.

While the scheme of magnetic logic was proposed almost a decade back, an experimental demonstration of such a device has not been shown yet. The proposal relies on ideal behaviors of the magnets, spin currents and the spin channels. While a metallic channel is an ideal choice for interconnect due to their low resistance, the small spin coherence length in metals will result in a loss of information before it is transferred from one magnet to another. Although a semiconducting channel supports a longer spin coherence length, the resistance mismatch between the ferromagnet and semiconductor makes the spin transfer efficiency very low. In addition, the superposition of spin currents coming from various inputs will highly depend on the interconnect length. Therefore, interconnects have to be designed precisely to obtain desired functionality, which may lead to scalability issues for a large number of logic array operation.

Bhowmik et al. demonstrated that the spin current generated using the SHE in a non-magnetic element such as Ta and Pt can perform the clocking function [28]. In their work, the magnetization state of an input magnet was used to control the final state of three nano-magnets as illustrated in Fig. 3(c). When the input magnet is set in the up (down) magnetization direction, the final state of the three bits stabilizes in down-up-down (up-down-up) configuration. The three nano-magnets which have dipole coupling between them change their state only when a charge current is passed through the underlying Ta layer, hence the clocking function.

The current-induced magnetization switching using SOT (refer to Section III for details on SOTs) occurs above a certain threshold value of current density. This property has been also used to build logic functions. A SOT device with two inputs in form of current acts as an AND gate. The final magnetization state of the magnet represents the output. The individual value of each current input is maintained below the threshold switching current of the device in order to achieve the AND functionality [29]. The direction and magnitude of the assist magnetic field was used to construct other logic functions such as OR, NAND, and NOR. For example, in order to implement an OR gate, the magnitude of assist field was increased such that the threshold switching current decreases below the individual input current values. This results in switching of the magnetization for all the input values except when both the inputs are zero, thereby emulating an OR gate. SOT devices which show a voltage control of magnetic anisotropy (VCMA) [30] have also been used to implement spin logic. The VCMA in these devices modulates the threshold switching current of the magnet [31] (Fig. 4). A multifunctional OR/AND gate was implemented using a single device depending on the initial magnetization states as shown in Fig. 4(c,d). The switching current and gate voltage acts as the two input parameters, while the output was measured using the anomalous Hall resistance ($R_{xy}$). These logic devices, however, have limited fan-out as the readout is carried using the anomalous Hall resistance. Again, device to device variation normally results in different threshold currents for different devices. The proposed device also requires an assist magnetic field which makes scalability difficult.

Spin waves which represent propagating disturbance in a magnetic material have also been proposed as a viable means to construct spin logic. The spin wave acts as information carrier and their phase can be varied by a magnetic field produced by current passing through a wave guide [32,33]. Spin waves from different sources add up constructively or destructively depending on input current values to realize Boolean functionalities. However, since the phase of spin wave is prone to disturbance from magnetic inhomogeneity and imperfections, it has been proposed to use the wave amplitude instead [34,35]. The nonreciprocity of spin wave's magnitude for opposite wave propagation direction can be exploited to implement a simple invertor or pass gates [34]. A larger value of nonreciprocity corresponds to a larger readout margin. It was shown that a Ta/Py bilayer system shows a giant nonreciprocity factor (the ratio of spin wave amplitude at positive and negative field) of ~ 14 (60 in the frequency domain) for the Ta thickness of 8.2 nm as shown in Fig. 5 [36]. The spin wave-based logic is still in a very preliminary stage. In comparison to charge currents, spin waves are a very weak information carrier. For sustaining reliable and long-distance transfer of information using spin waves requires specific magnetic materials. The demonstration of spin wave-based logic network is yet to be seen due to these limitations and requirements.

Magnetic logic devices utilizing skyrmions [37-39], which are topological spin states, have been proposed. Skyrmions can either be driven by the magnetic field or spin currents while they can be manipulated by dynamically modulating properties of the magnetic films using the electric field [40]. In one such scheme, this skyrmion behavior was utilized to simulate a skyrmion transistor. In the proposed transistor the skyrmions were driven by a spin current from one end of the spin channel to another. A gate in between the channel was used to annihilate the skyrmion by modulating the anisotropy of magnetic film under it, resulting in transistor "off" operation [41]. A hybrid structure based on skyrmion and DW has been put forward [42]. The DW and the skyrmion can be interconverted by designing magnetic channels of different widths. By designing specific nanostructures that duplicate, merge or annihilate skyrmions logic gates functionalities can be achieved [43]. While the skyrmion logic devices are promising, they are still in a conceptual and simulation stage. An experiment demonstration of their functionality and scalability is awaited with interest.

As we see in this Section, there have been multiple proposal and demonstration of spin logic devices and systems, however, these implementations still have a long way to go before they compete with CMOS logic. The normal propagation delay of any conventional CMOS logic gate is generally few ps. In comparison, the delay or speed of typical spin devices is limited to ns or GHz, respectively. The advantage offered by spintronics is, however, their non-volatile nature which can save a tremendous amount of power during data processing, as circuit blocks in execution pipeline but not being executed can be switched off. Interconnects is a big issue in all spin logic circuits. An ideal interconnect should be able to transfer spin information over long distances. However, in reality, spin interconnects made up of metals have relatively short spin coherence lengths making them unsuitable for spin logic circuits. In contrast, the interconnects in current silicon logic circuits can even run from one end of the chip to another and over many layers. This is due to the fact that an electric charge is a conserved quantity, but a spin is not, therefore it is not easy to transfer spin information for a long distance. Scalability is another important feature for any logic device. While a MTJ, which is the building block of spintronic memory is highly scalable even comparable to CMOS gates. The spin logic devices that have proposed and demonstrated till now fall short on this criterion. For example, the DW based logic devices that rely on the presence of DWs have dimension ranging from tens to hundreds of nanometers.

The majority of works on spin logic till now have been focused on single device demonstrations. The performance of these devices is still limited in terms of on/off ratio, speed, and scalability in their present form. For example, the on/off ratio of a thin-film-transistor panel is more than 100, which is regarded as a poor performance CMOS device. On the other hand, one of the best spin filter, an MgO tunnel junction has an on/off ratio less than 10. It is also essential to realize an interconnected network of spin logic devices to show significant advantage of a spin computing system over CMOS. The work on spin logic devices and systems is still in a very preliminary and exploratory stage. The viability of an all spin-logic network is yet to be demonstrated. A more viable alternative would be exploring hybrid CMOS-spin logic.

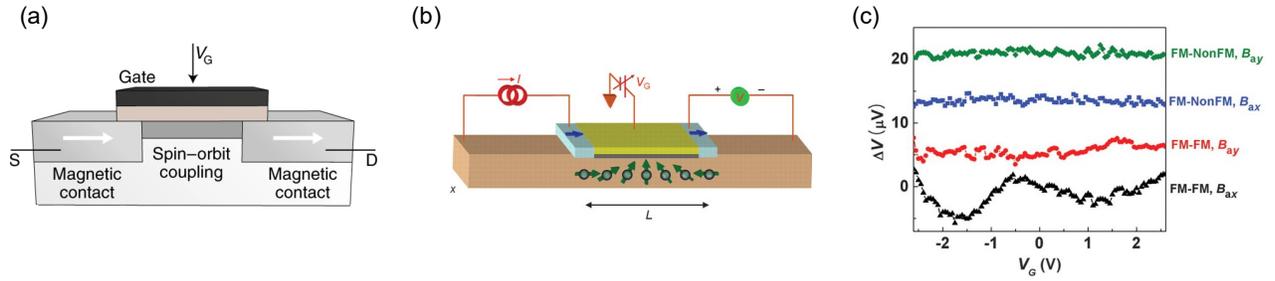

Fig. 1. (a) Schematic of the spin transistor. Reprinted figure with permission from [16]. [copyright statement]. (b) Gated spin valve device with the magnetic field applied along the channel direction. The injected electrons precess under the influence of the Rashba field. (c) Oscillatory conductance of the channel as a function of gate voltage ($V_G$) for different device types and applied external field directions. Oscillations are obtained only for devices when both the magnetic injector and detector are aligned along the channel direction by an external field. Reprinted figure with permission from [18]. [copyright statement].

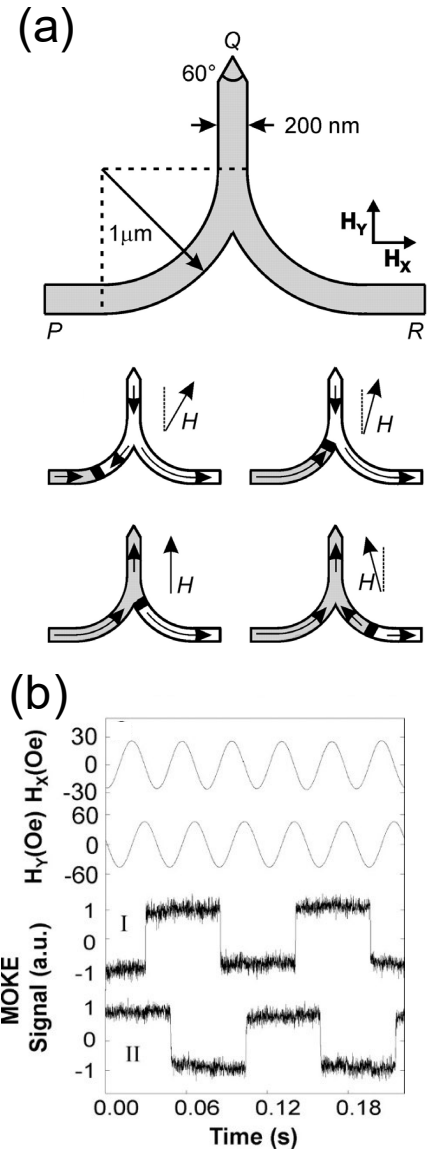

Fig. 2. (a) Schematic of a ferromagnetic NOT gate. The lower panel shows the operation principal of the device under the application of a rotating magnetic field. The thick line represents the domain wall. (b) The top panel shows the x and y component of the rotating magnetic field. The lower panel shows the MOKE traces at the input and output arm of the NOT gate. Reprinted figure with permission from [25]. [copyright statement].

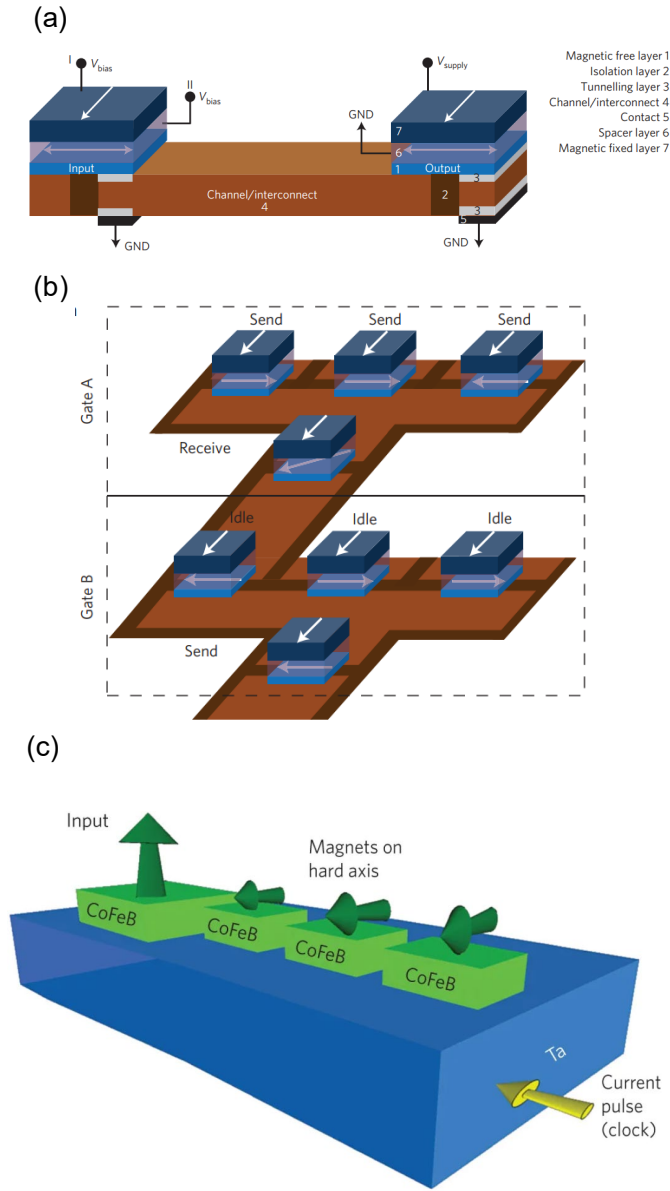

Fig. 3. (a) Illustration of an all-spin logic device. The output is held at a high energy state and is eventually switched using the spin current from the input bit. (b) Cascade gate design using all-spin logic devices. The state of the output is determined by the total spin current injected by the individual input bits. Reprinted figure with permission from [27]. [copyright statement]. (c) Schematic of a spin-torque clocking device. The state of the big magnet determines the configuration of the 3-bits when a spin current is applied using the spin Hall effect. Reprinted figure with permission from [28]. [copyright statement].

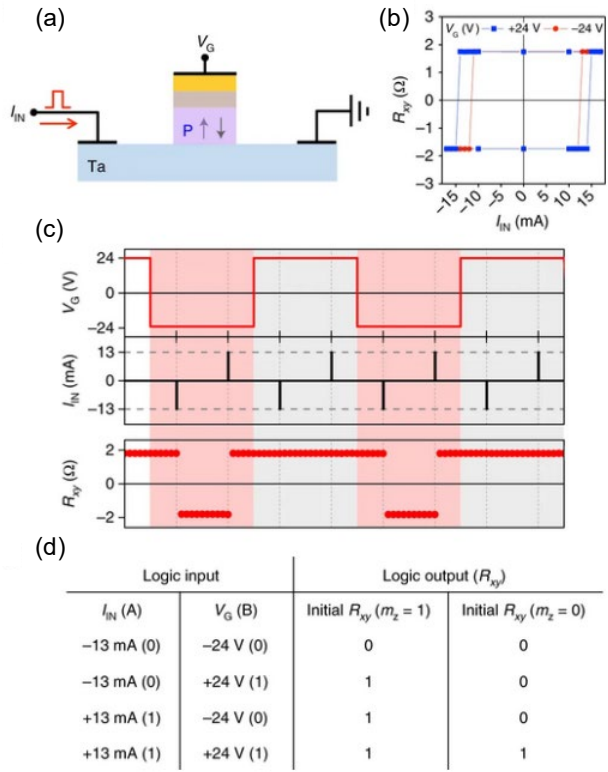

Fig. 4. (a) Device design and measurement configuration of a VCMA device. (b) The variation of switching current for two different gate voltages ($V_G$). (c) Output measured using the Hall resistance ($R_{xy}$) under the application of voltage and current of various polarities. (d) Input and output truth table for two different initial device states illustrates implementation of OR and AND logic functions. Reprinted figure with permission from [31]. [copyright statement].

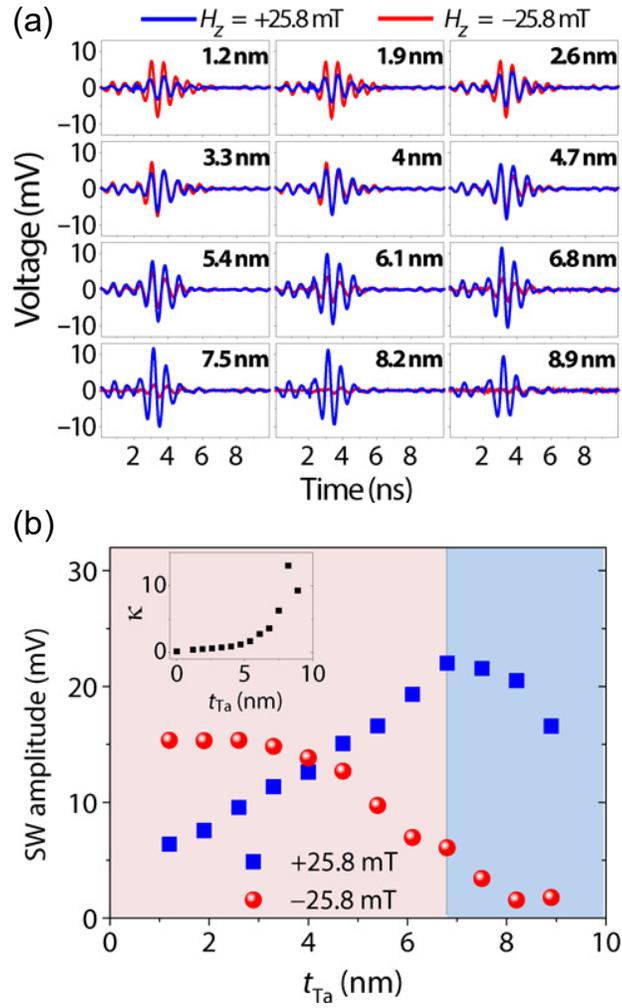

Fig. 5. (a) Spin wave packets measured for two opposite external magnetic field directions for Py/Ta devices with varying the Ta thickness ($t_{Ta}$). (b) Comparison of the spin wave amplitude for the devices. Inset shows the variation of nonreciprocity factor ($\kappa$). Reprinted figure with permission from [36]. [copyright statement].

## III. Spin devices for memories

The existing computing memory hierarchy has tremendous performance gaps in terms of the speed and density/cost which leaves abundant scopes for improvements. For example, the write speed of cache, main and eFLASH memory is around 5 ns, 30 ns and 0.1 ms, respectively [44,45]. Emerging memory technologies which have an intermediate speed between the cache and the main memory or that between the main memory and eFLASH are promising for bringing performance improvement in the future computing systems. In addition, future applications like Internet of Things (IoT) demand for fast computing at the edge in an energy-efficient manner. Spintronic in form of MRAM offers one such potential memory solution due to its non-volatility and high speed of operation [12,13,46].

The basic unit of a MRAM [47] is a MTJ [48] which stores digital information in the form of two bistable magnetization states of a thin magnetic layer. These magnetization states can be read-out using the resistance value of the MTJ. The writing of MTJ involves switching the magnetization from one stable state to another by crossing a high energy barrier between them. Ideally, it is desired to cross this energy barrier with a minimum input energy. At the same time, the energy barrier should not be too small so as to allow undesired magnetic switching due to the thermal energy from the surrounding. The MRAM research aims to develop device solutions that bring the balance between the conflicting requirements of a low switching energy and high thermal stability.

The first generation of MRAM also known as toggle MRAM used magnetic fields generated by the current in metal lines as shown in Fig. 6(a) to switch the MTJ. This writing method is both high energy consuming and non-scalable. The second generation MRAM which is based on STT (STT-MRAM) [49] operates on the principal of the transfer of angular momentum from a spin polarized electron to the atoms of the magnetic bit [5,49,50]. In a STT-MRAM illustrated in Fig. 6(b), the spin polarized current is generated by passing a charge current through a fixed magnetic layer. While offering tremendous advantage over the toggle MRAM, the STT-MRAM still requires a large write current and has a potential endurance issue due to the breakdown of the MgO barrier layer which separates the two magnetic elements (reference and free layer) of the MTJ [51]. In 2010, Miron et al. demonstrated that a spin current generated by passing charge currents through a non-magnetic heavy metal layer is capable of manipulating the magnetization of an adjacent magnet [52]. Their work formed the basis of the spin-orbit torque MRAM (SOT-MRAM) [52-55]. For STT operation, a nanopillar type MTJ is required to observe the current induced switching effect, which is very challenging in a typical academic institute, however, SOT devices even with the width of micro-size can still demonstrate the current induced effect due to a very thin film thickness involved in lateral current injection. This relaxed device patterning requirement attracted various academic institutes to SOT research. However, the device size of three-terminal SOT is larger than that of two-terminal STT, but is still half size of the modern static random access memory (SRAM). The SOT device also has an advantage over STT ones in terms of its speed. The incubation time in STT due to parallel alignments of the incoming spin and the magnetization of the free layer is absent in perpendicular SOT devices. The SOT geometry offers another advantage, the spin currents in SOT devices can be very large, as the electron interacts with the FM many times taking advantage of lateral scattering in the SOT device. In fact, Yoda et al. reported the switching efficiency of SOT is 3-4 times higher than that of STT using an in-plane MTJ device [56]. In the remaining part of this section we will discuss about the major developments in SOT-MRAM devices. At first, we begin by giving a brief introduction on SOTs.

A SOT device heterostructure typically consists of a SOT source adjacent to a magnetic layer which is the data storage unit [57,58] (see Fig. 6(c)). The SOT source is a material with large SOC, for example heavy metals (HM) such as Pt [57,59-61], Ta [61-64], W [65,66] etc. The magnetic layer is typically a metallic ferromagnet (FM) such as NiFe, CoFeB, etc. When a current is passed through the HM, spins accumulate at the interface of the HM and FM. The accumulated spins diffuse into the FM during which they transfer their angular momentum to the magnetic atoms. The overall result of this process is the manipulation of the FM magnetization and its eventual switching. The SOT research has been mostly focused on the understanding and exploiting the rich SOT physics to develop energy-efficient SOT heterostructures.

The spin accumulation and the subsequent spin current that is generated on passing a charge current through a SOT device is generally due to two physical mechanisms. The first mechanism is the SHE [67-73]. The SHE governs charge to spin conversion in a material with large SOC. Due to the SHE, charge currents carrying electrons with opposite spins are separated in two opposite directions resulting in a spin current ($J_S$). The direction of $J_S$ is orthogonal to both the charge current ($J_C$) direction and the spin polarization ($\sigma$) direction as illustrated in Fig. 7(a). Although predicted long back in 1971 [73], the SHE was not observed until 2004 [74] when it was measured using magneto-optical Kerr microscopy.

It should be noted that the spin accumulation due to the SHE is a result of charge currents flowing though the bulk of the HM. The spin separation in SHE arises either from the band structure of the SOC source (intrinsic SHE) [72] or the asymmetric spin-dependent scattering of the electron with the impurities in the SOC source (extrinsic SHE) [67,70]. The charge to spin conversion factor therefore depends on both the intrinsic and extrinsic factors, and it dictates the amount of spin current generated from a given charge current. The charge to spin conversion factor is called the spin Hall angle ($\theta_{SH}$). It should however be noted that in many of the reports the $\theta_{SH}$ is a combined representation of spin conversion factor from the SHE and other bulk spin current sources. Finally, a SOC source which exhibits the SHE also reciprocally converts a spin current to charge current through a process known as the inverse SHE (ISHE) [75,76].

The second process of the spin to charge conversion is through the interfacial Rashba-Edelstein effect also referred to as the Rashba effect [17,52,61,77,78]. In a heterostructrue with broken inversion symmetry (e.g. SOT devices) an electric field exists at the interface of two different layers. For example, in a SOT device an electric field ($E$) exists at the interface of HM and FM. When an electron flows through this interface, the electron experiences an effective magnetic field in a direction given by $E \times p$, where $p$ is the electron's momentum (see Fig. 7(b)). Under the effect of this relativistic magnetic field, the electrons at the interface are polarized in the direction $E \times p$. These spin polarized electrons diffuse into the adjacent magnet and manipulate its magnetization similar to the SHE. However, unlike SHE, the Rashba effect is an interfacial effect and does not depend on the current flowing through the bulk of the HM. In reality, both the SHE and Rashba effect are present in a typical device and its relative contribution is not easy to be distinguished, even though thickness dependence studies are often utilized for this purpose. We encourage the readers to refer to the focused reviews on SHE and interfacial SOC effect for a detailed physical insight on these phenomena [68,69,79,80].

While most of the experimental works on evaluation of the SHE, Rashba effect, and SOTs involve their indirect measurements using an adjacent magnet, the first observation of SHE was a direct imaging in semiconductors GaAs and InGaAs using magneto-optical Kerr microscopy [74]. A similar examination of spin accumulation in metallic HM systems with Kerr microscopy has yielded controversial results. With some groups claimed to have observed the Kerr rotation in the presence of accumulated spins [81,82], others suggest the SHE signal to be too weak to result in any Kerr rotation for these systems [83,84]. It was suggested that the observed signal probably arises from the change in reflectivity of the metal due to heating. With the debate still open on the validity of MOKE to visualize the SHE in metals, an alternate method of photoconductance has been employed recently to visualize the current-induced spin accumulation [85,86]. In a photoconductance measurement, a circularly polarized laser is shone on the channel which results in a voltage difference across it. When a current is also passed through the channel, it was observed that the generated voltage has helicity dependence due to magnetic circular dichroism. Figure 8 shows the spatially resolved photovoltage map for $Bi_2Se_3$ (topological insulator) and Pt. The helicity dependent photovoltage polarity is reversed on changing the current direction in the channel. This observation is in line with the reversal of the spin polarization direction due to the change in electron's momentum.

We will discuss the recent advances in the development of SOT devices. The sub-sections have been divided in terms of the approaches used for these developments.

### A. Novel SOC source materials

The most straightforward approach to develop devices with a high SOT efficiency is by replacing the underlying SOT/SOC source. Conventional SOT devices incorporate a HM layer as the source of SOTs. Typically, HMs such as Pt, Ta, W, and Hf are used in these devices which have $\theta_{SH}$ in the range of 0.05 to 0.5 [57,62,65,87,88] which makes these materials very efficient in charge to spin conversion and subsequent magnetization switching. Alloys involving two different heavy metals [89,90] (e.g. AuPt), heavy metal with a low SOC material like Cu, etc. have also been shown to exhibit a large value of $\theta_{SH}$ [91-97]. However, there has recently been an interest in exotic and novel materials such as topological insulators (TIs) [98-107] and Weyl semimetal [108-121] due to their large charge to spin conversion efficiency.

TIs are materials with an insulating bulk and conducting surfaces (topological surface states, TSS) [122,123]. Interestingly, the TSS exhibit spin-momentum locking in which the spin polarization direction of an electron on the TSS is fixed with respect to its momentum. This results in spin accumulation at TSS similar to the SHE and the Rashba effect observed in HMs. In spintronics, TI research has been pursued both for their interesting physics and practical applications in SOTs. In 2014, it was shown that current induced spins in TIs $(Bi_{0.5}Sb_{0.5})_2Te_3$ can switch a magnetic element [101]. However, this experiment was performed at 5 K and a magnetic doped TI was used as the switching element instead of a metallic FM. The first room temperature magnetization switching using TI was performed on a ferrimagnet (Fig. 9(a)) [107] and on a metallic FM [104]. On the latter, the switching was observed using MOKE microscopy, and a TI thickness dependence study was also performed. The SOT efficiency in TI ($\theta_{TI}$) as a function of thickness can be divided in three regions (Fig. 9(b)) depending on the dominating source of SOTs for these thicknesses. While the spin-torque from bulk states (current flowing through the bulk) dominates the spin accumulation in Region I, the large enhancement of $\theta_{TI}$ is attributed to the TSS in Region-III (smaller thickness). Pan et al. have mapped the spin texture of the TSS using a very simple yet effective electrical measurement technique of bilinear magnetoresistance (BMR) [124]. Both in-plane and out-of-plane spin textures were mapped using this method. Figure 9(c) shows a 30° canting angle of the spin with respect to the film plane as measured using the BMR probe.

Another exotic material system for SOT research is the Weyl semimetals [125]. They have been predicted to have large spin splitting i.e. Edelstein effect due to their non-trivial band structure [115]. Current Weyl semimetal SOT researches are largely focused on $WTe_2$. The crystal structure of a Weyl semimetal, WTe$_2$ has only one mirror plane and does not contain a two-fold rotational invariance (Fig. 10(a)). Therefore, the current-induced spin accumulation response in WTe$_2$ is anisotropic. In fact, when the current is passed through a low-symmetry axis (a-axis), a sizable out-of-plane spin accumulation is detected [118,119]. This behavior was evident from the asymmetric spin-torque ferromagnetic resonance (ST-FMR) spectra when the magnetic field was applied in two opposite directions (Fig. 10(b)). This unique property of WTe$_2$ makes it ideal for use in SOT devices with perpendicular magnetic anisotropy, while a few other field-free SOT switching schemes were proposed [126-130]. Current-induced magnetization switching using WTe$_2$ has been recently demonstrated, shown in Fig. 10(c,d), and an in-plane $\theta_{SH}$, which increases up to a maximum value of 0.8 with increasing the WTe$_2$ thickness has been reported [119]. Interestingly, the required power to switch the magnetization was 19 times smaller in WTe$_2$/Py than that of Bi$_2$Se$_3$/Py and 350 times smaller than that of Pt/Py due to a high efficiency and low resistivity of WTe$_2$.

Apart from TIs and Weyl semimetals, several other 2D materials such as MoS$_2$ and WSe$_2$ have been also shown to have a moderate charge to spin conversion efficiency [131,132]. 2DEGs at the interface of LaAlO$_3$ and SrTiO$_3$ also support a giant charge-induced spin accumulation [133-137]. While the above mentioned exotic materials have been promising in terms of their charge to spin conversion properties, the biggest hindrance towards their practical applications is the difficulty in their fabrication for a large and uniform area as well as their large resistivity leading to a current shunting issue. Single crystal TIs, semimetals, and other layered materials in majority of the reports discussed above have been either exfoliated or fabricated using sophisticate molecular beam epitaxy (MBE). Only very recently sputter deposited TIs have been shown to produce large SOTs capable of switching an adjacent magnet [106,138]. While sputter deposited exotic materials are expected to be more technologically relevant, it is still not clear whether topological features are present and what the role is in the structure with small crystalline clusters and non-stoichiometric inhomogeneous composition. Future research effort should focus on fabrication of these novel materials in a fast, efficient and reliable ways.

**B. Engineering the magnetic layer**

While the SOTs nominally arise from the SOC source such as HMs, the magnetic layer can also serve as a secondary source and modulator of the SOTs. Among the different types of magnetic materials, ferromagnets which have positive exchange interaction between the individual atoms or layers have been widely explored for SOTs. However, there are also another class of magnetic materials such as antiferromagnets (AFMs) and ferrimagnets (FIMs). These materials typically consist of two different atomic sub-lattices that prefer to align their spins opposite to each other due to the negative exchange interaction between the two sub-lattices. While AFMs have net zero magnetization due to equal and opposite magnetization of the constituent sub-lattices [139,140], FIMs have non-zero magnetization due to unequal magnetization of the constituent elements. The major advantage of AFMs and FIMs over FMs is their robustness against external magnetic fields. This makes them thermally very stable resulting in a longer retention time compared to FMs. The thermal stability in particular is a very important factor for scaling down the spintronics devices in the nanometer regime.

In 2016 it was first shown by passing a current through an AFM with locally broken inversion symmetry such as CuMnAs [141], the antiferromagnetic domains can be switched [142]. The switching mechanism in CuMnAs is illustrated in Fig. 11(a). Current-induced local magnetic field is generated around individual Mn atoms due to the inverse spin galvanic effect [143,144] which requires broken inversion symmetry in the given system. While the CuMnAs crystal as a whole has inversion symmetry, the local environment of two sub-lattice formed around Mn atoms has broken inversion symmetry. This results in staggered magnetic field of opposite polarity around these atoms as shown in Fig. 11(a). The staggered magnetic field results in switching of the AFM Néel vector as a whole. The switching of the Néel vector was detected using the anisotropic magnetoresistance (AMR). Figure 11(b) shows the varying AMR signal on application of current pulses.

Later, AFM switching was also demonstrated in $Mn_2Au$ [145,146]. Since a single current pulse moves the AFM domain only very slightly, multilevel memory cells were demonstrated with CuMnAs [147]. The ultrafast dynamic of AFMs allow their switching with picosecond current pulses [148]. The AFM domains have been imaged using x-ray magnetic linear dichroism-photoemission electron microscopy (XMLD-PEEM) as shown in Fig. 12(a-c) [149]. Chen et al. have recently shown that the magnetic anisotropy of Néel vector in $Mn_2Au$ deposited on a ferroelectric substrate, PMN-Pt, can be switched between two orthogonal directions [150]. This results in a ratchet like switching behavior (see Fig. 12(d)). Apart from inducing switching, it has been also proposed that the staggered relativistic SOT fields can result in a very high domain wall velocity in these AFMs [151]. Other than AFMs with broken inversion symmetry, there have also been demonstration of Néel vector switching in insulating AFMs such as NiO [152-154]. In these works, the SOTs are generated from an adjacent heavy metal instead from the AFMs themselves.

While the AFMs present an exciting and stable system for spintronic applications, the detection of their magnetization state remains challenging. Since the AFMs are read electrically using a small value of AMR rather than TMR, they are incompatible with a magnetic tunnel junction (MTJ) which is the standard readout device in the MRAM industry. For example, a TMR value of 150-200% is required for a readout speed of 5-20 ns. In addition, a recent work by Chiang et al. suggests that the AMR measurements in AFMs are complicated by the normal Seeback effect and resistive switching [155]. On the other hand, synthetic antiferromagnets (SAF) and FIMs offer viable alternative to both the AFMs and the FMs. Not only they are unperturbed against external magnetic field similar to the AFMs, their magnetization can be also detected easily by anomalous Hall resistance ($R_{AHE}$) and TMR similar to the FMs.

Synthetic antiferromagnetism arises from the interlayer Ruderman–Kittel–Kasuya–Yosida (RKKY) coupling between two ferromagnetic layers separated by a spacer such as Ru [156-160]. Compared to AFMs such as CuMnAs, SAF are easier to fabricate and do not require special crystalline substrate. Figure 13(a) shows a measurement of a very large domain wall velocity (750 m s$^{-1}$) obtained in a SAF formed by Co/Ni/Co multilayers through a Ru spacer [161]. The negative exchange coupling results in this high domain velocity which increases with the amount of compensation (Fig. 13(b)) between the two coupled layers. It was also found that the SOT switching efficiency in a completely compensated SAF made with Co/Pd FM layers was significantly higher compared to the FMs [162]. Recently, it was shown that the SOT for the AFM coupling case is ~15 time larger compared than without the AFM coupling in a Pt/Co/Ir based SAF system [163]. This report suggests that the interface induced phenomena apart from the negative exchange torque [161,164] are possibly responsible for such a large SOT efficiency in the SAF.

The relatively weak exchange coupling in the SAFs leaves space for the exploitation of the FIMs. Rare-earth transition-metal (RE-TM) ferrimagnets [165,166] have been the material of choice in magneto-optical disks previously and FIM SOT devices recently. The magnetization in these RE-TM FIMs can be tuned by varying the relative composition of the RE and TE components or by adjusting the ambient temperature. The tunable characteristic of RE-TM FIMs provides a fertile ground for studying SOT in both the AFM and FM regimes. In CoGd, CoTb and GdFeCo based FIMs, it has been shown that the SOT efficiency or the SOT effective fields increase near the compensation point i.e. when the magnetization of the FIM is minimum [164,167-172]. However, the amplification of SOT near compensation cannot be fully explained by the reduced magnetization. It has been proposed that apart from reduction of the magnetization near compensation, the drastic amplification of SOTs in compensated ferrimagnets has been attributed to the enhanced negative exchange torque [162-164]. The high efficiency of SOTs in the RE-TM FIMs and the fact that these materials derive perpendicular anisotropy from the bulk, helps in developing devices with thick and thermally stable magnetic layer. Roschewsky et al. have shown current-induced magnetization switching of 30 nm thick GdFeCo layer which has thermal stability of ~ 100 $k_B$T, where $k_B$ is the Boltzmann constant and T is the temperature [172].

The faster intrinsic dynamics due to the negative exchange coupling also results in ultrafast switching in the FIMs. Figure 13(c) compares the switching energy and switching duration of the FIMs to that with the FMs [173] from which it can be seen that the FIMs are around 1-2 orders of magnitude better than FMs on both these parameters. The ultrafast dynamics in FIMs [174,175] also leads to a very high current-induced DW velocity up to 5.7 km/s [173] when compared to the FMs or even the SAFs (Fig. 13(d)) [176,177]. The high DW velocity finds applications in DW motion based magnetic devices. It has also been reported that the FIMs and AFMs support longer spin coherence lengths [178,179]. Figure 14(a) illustrates the alternating spin alignment in a ferrimagnet that assists spin information transfer over a longer distance. Due to opposite exchange fields in alternating sublattices, spin dephasing is partially compensated. The SOTs therefore act over a larger thickness in the FIMs as shown in Fig. 14(b) showing bulk-like SOT [178] when compared to their limited penetration in (< 1.2 nm) in the FMs [180,181]. A disadvantage of RE-TM FIM is that the RE element is prone to oxidation. In MRAM fabrication steps in which the die is exposed to different temperatures and chemicals, preventing oxidation of FIMs can be challenging. The temperature sensitivity of the magnetization of FIM and related properties should be considered while designing memories based on them. For example, the temperature sensitivity of FIM could be a critical issue involving varying temperature environment such as automotive applications. Considering that FIM has been successfully commercialized for magneto-optical disks, the above oxidation and temperature issues could be also addressed in magnetic memory applications.

While in this section we have mainly discussed the research works concerning the antiferromagnetically coupled materials, there are analogous efforts to exploit other interesting magnetic systems such as multilayers [182,183], magnetic insulators [184,185], etc. Apart from a faster speed and lower power consumption, the integration of new magnetic systems in the MTJ based MRAM architecture is a criterion that is important to meet for real applications.

## C. Engineering the SOT heterostructure stack

The simplest SOT device consists of a SOC source neighboring a magnet. In this section we will discuss about alternative SOT heterostructures that enable an enhanced efficiency. In 2014, Woo et al. demonstrated that by sandwiching the FM between SOC sources with opposite $\theta_{SH}$, the overall SOT effective fields in the device can be enhanced [186]. In their work, a thin Co layer was inserted between Pt and Ta which are the materials with opposite $\theta_{SH}$. The maximum effective $\theta_{SH}$ of the device was found to be around 0.35 for a Ta and Pt thickness of 4 and 3 nm, respectively. This value is larger than the individual $\theta_{SH}$ of either Pt or Ta. Later in a detailed report, Yu et al. compared the effective field among different material combinations (e.g. Pt, Cu, MgO, Ir, Ta, W and Hf) with Pt as shown in Fig. 15(a) [187]. It was found that a combination of Pt and W results in the largest effective $\theta_{SH}$ of 0.45. Additionally, it was suggested that the thicknesses of the two

HM layers are critical for obtaining the largest SOT efficiency, as it dictates the current shunting in the individual layer. It has been also shown that a bilayer of two heavy metals (Pt and Ta) can be used to continuously tune both the magnitude and direction of SOTs [188].

The interface between the HM and FM is critical in determining the magnitude of SOTs in a given device. The quality of this interface decides the amount of spin memory loss and spin transparency, hence the magnitude of spin current experienced by the FM [189-191]. Due to the above reasons, the experimentally obtained SOT strengths (e.g. $\theta_{SH}$) are effective values including various bulk and interface effects. It was shown that by dusting a thin layer (~ 0.5 nm) of Hf at the Pt/CoFeB interface, the spin-torque efficiency could be enhanced up to 0.12, which is nearly double compared to the Pt/Py system [192]. An accompanied reduction of the magnetic damping, $\alpha = 0.012$, was also reported because of spin pumping suppression. The above approach was used to achieve a low switching current density of $5.4 \times 10^6$ A/cm$^2$ in an in-plane MTJ which had Hf dusting on both interfaces of the free layer [193]. An interfacial Ti layer (~ 1 nm) between Pt and CoFeB also helps in reducing the switching current density three times [194]. Recently, it has been suggested that a lower value of interfacial SOC at the HM/FM interface results in a lower attenuation of the spin currents by decreasing the spin memory loss [195].

Capping the HM/FM stack with Ru layer has been shown to result in an enhanced value of SOTs [181]. This enhancement is attributed to the spin current absorption in the Ru layer as illustrated in the schematic of Fig. 15(b). The SOT effective field increases by a factor of three for a Ru thickness of 0.6 nm when compared to a Pt/Co/Ni/Co device without any Ru layer on the top. In another variation of the SOT devices, it has been demonstrated that an in-plane magnetized FM layer (Fig. 15(c)) can perform charge to spin conversion similar to a HM [196-198]. The resulting spin current in this heterostructure is capable of switching a perpendicular CoFeB layer without any assist in-plane magnetic field (see Fig. 15(d)) due to a slight out-of-plane alignment of the generated spins. However, in this heterostructure the in-plane magnet should be immune against the external magnetic field. Therefore, an in-plane magnet with a high coercively is required. Moreover, repeated switching events can perturb the magnetization direction of the in-plane magnet and thus affecting the long term stability of the memory device. Inadvertent switching of the in-plane magnet can result in opposite sense of switching loop and can lead to unreliable memory operations. We would like to bring to the reader's attention that apart from the aforementioned works on engineering SOT heterostructures, there is a dedicated research effort on engineering SOT heterostructure for field-free SOT switching of the perpendicular magnetization, which is of significant importance for applications, the details of which can be found in a recent review [53]. Even though many different schemes were proposed for field-free switching, there is no clear avenue which can satisfy requirements of MTJ integration and large wafer size scaling with a homogeneous device to device uniformity, and further research is required to address this important issue.

**D. SOT modulation by oxygen incorporation**

Oxygen plays an important role in determining many magnetic properties. For example, the incorporation of oxygen in metallic FMs such as Fe, Ni and Co, induces negative exchange interaction resulting in the formation of AFMs such as $Fe_2O_3$, NiO and CoO. For applications in nanodevices, a strong perpendicular magnetic anisotropy (PMA) which is vital for device scalability is derived from the orbital hybridization of magnetic atoms with the oxygen at the interface [199,200]. As a result, the PMA has a strong correlation with the amount of oxygen at the magnetic interfaces. In fact, it was shown that by migrating the interfacial oxygen using the electric field, the PMA can be significantly altered in a non-volatile and reversibly way [201-203].

In the field of SOTs, it was found that the oxide capping layer plays a role in determining the magnitude of SOTs for a given device [204]. Both the field-like and damping-like torques were found to be 10 and 6 times larger, respectively, for a MgO capped Hf/CoFeB device compared to a TaO$_x$ capped one. The different interfacial electric field at the oxide interface and the resulting Rashba torque was attributed to this variation [87]. This suggests that not only the bottom HM/FM interface, but also the top FM/capping interface should be considered to estimate the overall SOT effect in a given structure.

Incorporation of oxygen in the magnetic layer is another way of manipulating the SOTs. It was first shown that when oxygen is dynamically introduced in the Co layer of a Pt/Co/GdO$_x$ device by the application of a gate voltage, the resultant SOT was an order of magnitude larger compared to the unmodified device [205]. While an enhancement of SOT on Co oxidation is expected due to the reduced Co magnetization, the observed amplification of SOT in this work was found to be disproportionately larger compared to the amount of reduction in the magnetization. This hints towards a role of oxygen in modifying the interfacial SOTs. A similar enhancement of SOTs was also reported on a HfO$_x$ capped device in which the gate voltage was applied using ionic liquid [206]. Hasegawa et al. recently showed that on introducing an oxidized Co layer at the Pt/Co interface a 4- and 10-fold enhancement of the longitudinal SOT effective field ($H_L$) and transverse SOT effective field ($H_T$), respectively, was achieved [207].

Apart from modulating the magnitude of SOTs, an oxidized Co or CoFeB layer on a thin Pt layer results in a reversal of the overall spin accumulation direction or the effective $\theta_{SH}$ polarity [208,209]. It was shown by Qiu et al. that in a series of Pt/CoFeB/MgO/SiO$_2$ devices with varying the SiO$_2$ capping thickness, devices with thin SiO$_2$ capping ($\leq$ 1.5 nm) have opposite SOT effective fields compared to the ones with thicker SiO$_2$ capping [208]. Interestingly, the SiO$_2$ thickness of 1.5 nm corresponds to the native oxide thickness of $SiO_2$ when it is exposed to the air. This opposite direction of SOT fields resulted in an opposite current induced switching polarity as measured by the anomalous Hall resistance ($R_H$) in the Fig. 16(a). It was found that the thickness of the $SiO_2$ capping determined the amount of oxygen in the magnetic layer, hence the polarity of SOTs. Recently, it was shown that this SOT polarity control can be achieved dynamically and reversibly in a single Pt/Co/GdO$_x$ device using electric field assisted oxygen migration [209] as illustrate in Fig. 16(b), in which GdO$_x$ works as an oxygen reservoir sending and receiving oxygen ions depending on a negative and positive top gate bias voltage, respectively. The SOT polarity reversal has been attributed to competition between the bulk SHE and interfacial Rashba effect. An oxidized Pt/Co or CoFeB interface has a larger Rashba torque with an opposite polarity compared to a spin Hall torque, therefore the device has a negative SOT polarity. The critical level of oxidation of the Co layer is found to be ~30 to 40% in order to observe a SOT sign change, however over-oxidation (> 50%) can cause an irreversible formation of CoO, which cannot be reduced back to Co by applying a gate bias. In addition, the oxygen at the interface was fine-tuned with the electric field to program the SOT device with a range of effective spin Hall angle as shown in Fig. 16(c). Such a sign change behavior with oxygen cannot be understood by the spin Hall physics. A recent work revealed that the experimentally reported oxygen-induced sign reversal of the SOT in Pt/Co bilayers is due to the significant reduction of the majority-spin orbital moment accumulation on the interfacial HM atoms [210]. One of the drawbacks of oxygen migration based spintronic devices is their slow speed of modulation. Unlike electrons, the oxygen ions migrate relatively slowly requiring between few ms to tens of seconds to change the state of given device. The optimization of the gate oxide thickness and material is essential to improve the performance of these devices. For example, recently yttria-stablized zirconia was used as an gate oxide to achieve anisotropy modulation in ms, which is 100 times faster than any of the previously demonstrated magneto-ionic devices [211]. Due to a slow speed, these devices are not suitable for near core memories, but can be exploited for flash replacement and field-programmable gate array (FPGA). In addition to the slow speed, the constant flux of oxygen ions inside these devices may result in a breakdown of the oxide or even the magnet after many cycles of operation. Since the process of oxygen migration is stochastic, there is cycle to cycle variability in the device operation. All the above concerns of oxygen migration based spin devices need to be addressed before their practical applications, which share similar fundamental challenges with resistive random-access memory (RRAM) due to the ionic migration nature.

The oxidation of the SOC source or the HM has been used in few works to modulate the SOTs. A large $\theta_{SH}$ of $-0.5$ was reported when 12 % of oxygen was incorporated in tungsten [212]. The $\theta_{SH}$ remains relatively insensitive with increasing the oxygen content (Fig. 17(a)) even though the material properties of W change with oxidation. The giant $\theta_{SH}$ was attributed to β-phase stabilization of W and an increase in the interfacial SOTs. However, the high resistivity of W which further increases with oxygen incorporation will results in large power consumption when these devices are used in memories. Recently it was reported that an oxidized Pt, which is an insulator, shows $\theta_{SH}$ comparable to that of normal Pt, even capable of inducing efficient magnetization switching [213,214]. While W and Pt are heavy metals with large SOC, a similar enhancement of SOTs on both surface and bulk oxidation of Cu was observed in a series of studies [215,216]. Fig. 17(b) shows the ST-FMR measurement results for different degree of Cu surface oxidation. A NiFe/Cu sample with a naturally oxidized Cu surface was found to have a larger spin-torque efficiency as evident from the symmetric ST-FMR component when compared to the SiO$_2$ capped un-oxidized Cu sample [215]. The SOTs in oxidized Cu were attributed to intrinsic Berry curvature and modification of orbital hybridization [216]. Overall, we see that oxygen in different layers and interfaces of a SOT device plays a vital role in enhancing and manipulating SOTs. Future research efforts should aim toward a more dynamic modulation of this oxygen content in a single device using gate biasing for applications.

Apart from the spin memory devices discussed in this section which are mainly based on switching of a magnetic element, there have been parallel efforts to enable magnetic memories using alternate schemes. One such methodology involves skyrmions on race tracks [217]. A race track is a magnetic channel that can be used to store memory bits in form of magnetic domains [218]. These bits or magnetic domains can be moved with the help of spin currents. In the skyrmion race track memory, these magnetic domains are replaced by skyrmions which have greater topological protection. In addition, the small size of skyrmion and the ease with which can be moved with a current, holds promise towards developing a denser and low-energy storage. The spin current to move the skyrmions can be applied either using STT or SOT. However, there are many challenges that are currently being addressed in the field of skyrmion devices before a reliable memory can be realized. The corresponding research work primarily focuses on stabilization of skyrmion, their energy-efficient and linear movement on a magnetic track and performing low-noise reading. More details of these works can be found in focused reviews [38,39].

Spin devices are one of the most promising non-volatile memory candidates. Spintronics memories have evolved from a field-switching toggle MRAM to current-switching STT-MRAM and then recent active research topic of SOT-MRAM. There are still a few major challenges that need to be addressed before the practical implementation of SOT-MRAMs. First of all, the proposed material system and device structure should be compatible with the MTJ read-out scheme for fast read operations and a large size wafer scaling with a good uniformity should be guaranteed. External magnetic field-free switching of SOT devices without involving a complicated structure is another requirement for an adoption of SOT-MRAMs. Other challenges include a further reduction of the writing power and the switching time for static random access memory (SRAM) replacement. Among the SOC candidates it is desirable to have a good balance between the spin Hall efficiency and charge conductivity. While the topological insulator such as $Bi_2Se_3$ and heavy metal such as W offer a large spin Hall angle, their large resistivity values will lead to large power consumption when used in memories. The switching endurance and read/write error rate evaluation also need to be carried out to meet the application requirements.

Among the explored physical mechanisms for non-volatile memory applications, spin memories hold competitive advantages in most of the performance parameters. The area of spintronic RAM is around 10-30 $F^2$ which is ~ 5 − 10 times smaller than semiconductor SRAM and is comparable to other non-volatile candidates such as the RRAM, phase change memories (PCRAM) and ferroelectric RAM (FeRAM) [44]. The area per cell of RRAM can, however, be reduced by 3D architecture. The write voltage of STT-RAM of ~ 1 − 1.5 V is the lowest among all the memories and is comparable to the SRAMs and DRAMs. For example, the write voltage for RRAM and PCRAM is greater than 3 V while that of FLASH memory is easily greater than 10 V. In terms of the write energy, all the non-volatile memories are more power consuming (0.1 − 100 pJ) compared to the SRAM and DRAMs (1 − 10 fJ). However, STT-RAM falls at the lower end of this spectrum, consuming only 1 pJ during the write operation. In comparison, the eFLASH easily consumes around 100 pJ during writing. The lowest power of ~ 0.1 pJ is consumed by FeRAM. With adoption of SOTs in the MRAM, the writing power can be even reduced further. The reading and writing speed of MRAM is also relatively high (~ a few to tens of ns). The writing time is in fact 5-6 order lower than the eFLASH. The endurance of STT-MRAM is predicted to be around $10^{15}$ cycles. RRAM which is based on movement of ions and PCRAM which involves thermal treatment in every write cycle, suffer from a relatively low endurance of $10^7$ to $10^{12}$ cycles due to the nature of their operations. The endurance of MRAMs can be further improved by the SOT architecture which separates the read and write path.

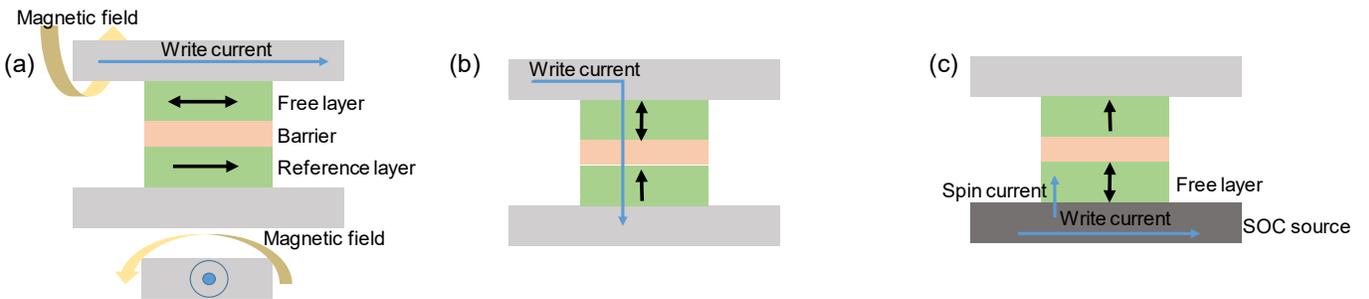

Fig. 6. Writing mechanism for a (a) Toggle-, (b) STT-, and (c) SOT-MRAM cell.

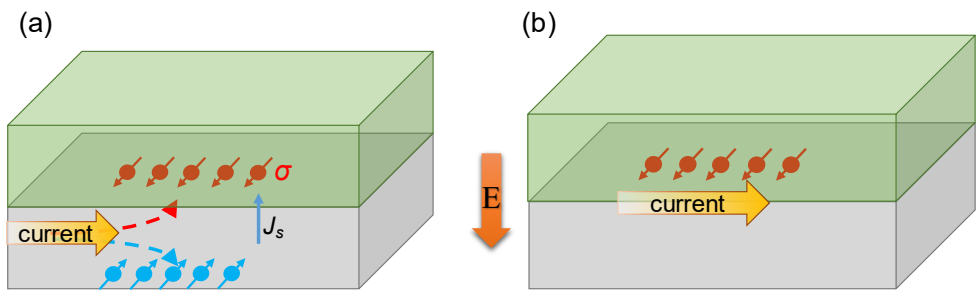

Fig. 7. Mechanisms of current-induced spin accumulation due to (a) Spin Hall effect and (b) Rashba effect. σ is the spin polarization direction, and $J_S$ is the spin current.

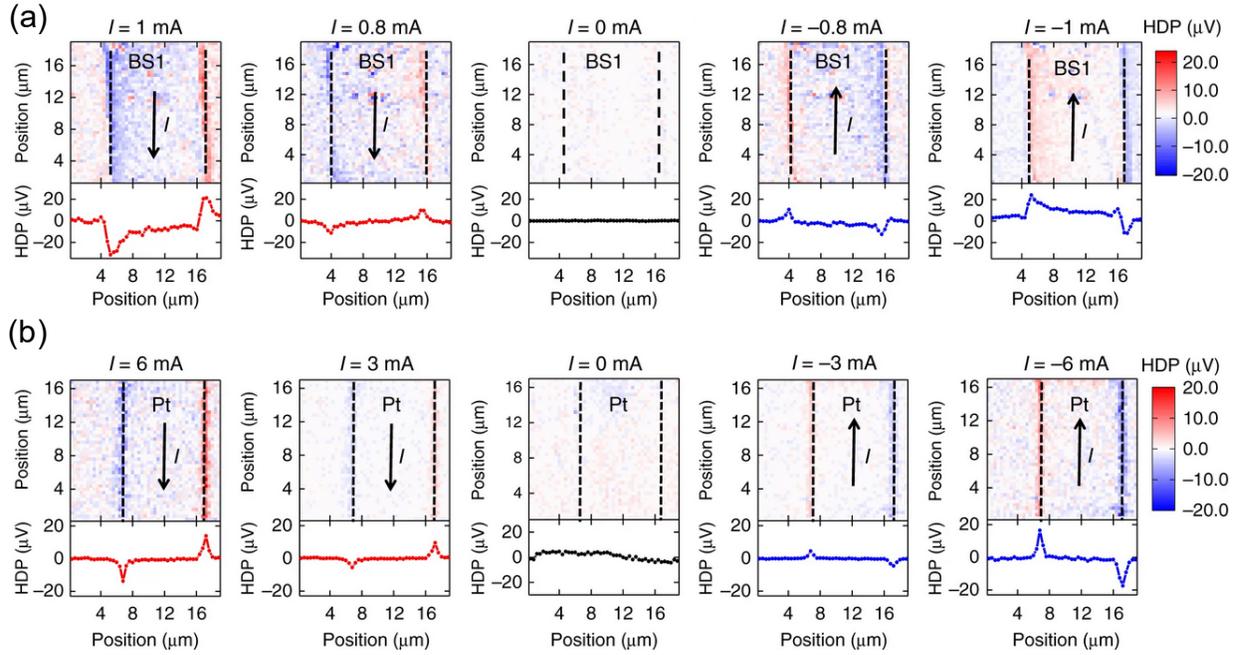

Fig. 8. Current-induced spin accumulation visualized using helicity dependent photoconductance (HDP) map for (a) $Bi_2Se_3$ and (b) Pt. Reprinted figure with permission from [86]. [copyright statement]. Note that there is out of plane spin accumulation on the top surface in $Bi_2Se_3$ due to hexagonal warping, which changes the sign with the current direction ($I$).

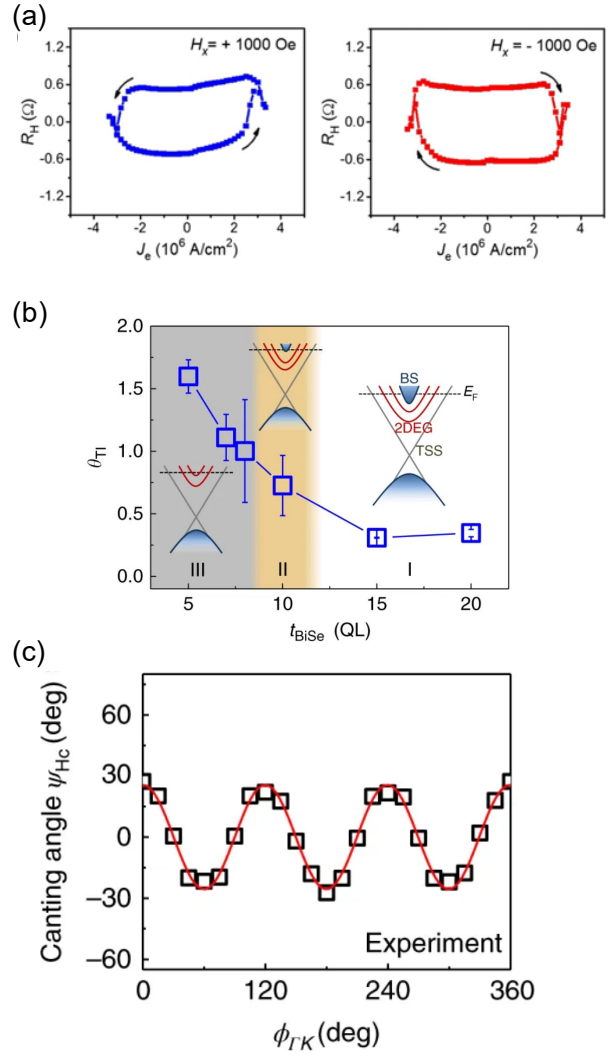

Fig. 9. (a) Current-induced switching of perpendicularly magnetized CoTb using a topological insulator, $Bi_2Se_3$, at room temperature. Reprinted figure with permission from [107]. [copyright statement]. (b) Dependence of room temperature SOT efficiency ($\theta_{TI}$) of topological insulator as a function of its thickness. Reprinted figure with permission from [104]. [copyright statement]. (c) Out-of-plane spin canting angle as a function of the angle between the current and $\overline{\Gamma}\overline{K}$ axis of the topological insulator. Reprinted figure with permission from [124]. [copyright statement].

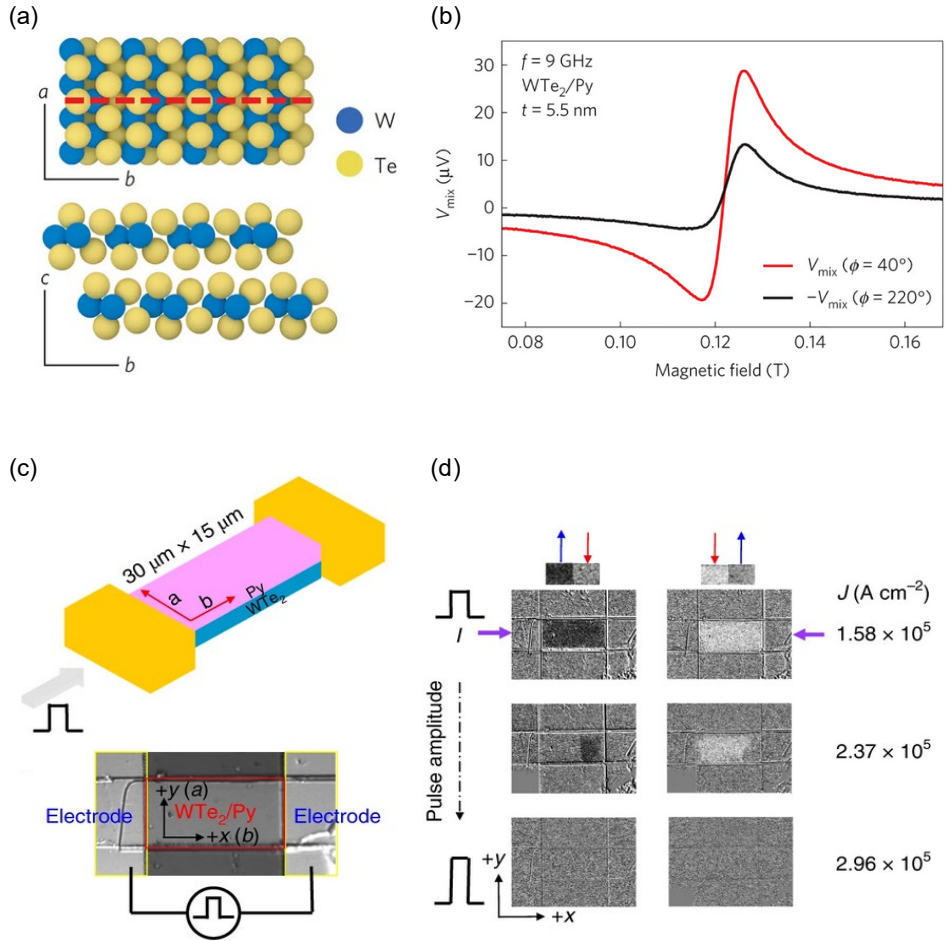

Fig. 10. (a) Crystal structure of WTe$_2$ near its surface. The crystal is symmetric with respect to the red dashed line in the *ab* plane but not symmetric with respect to the *ac* plane. (b) Measured ST-FMR signal for a WTe$_2$/Py device for two opposite external field directions. Reprinted figure with permission from [118]. [copyright statement]. (c) Schematic of a WTe$_2$/Py device and its device image. (d) MOKE images during current-induced switching measurement for the in-plane magnetized WTe$_2$/Py device. Reprinted figure with permission from [119]. [copyright statement].

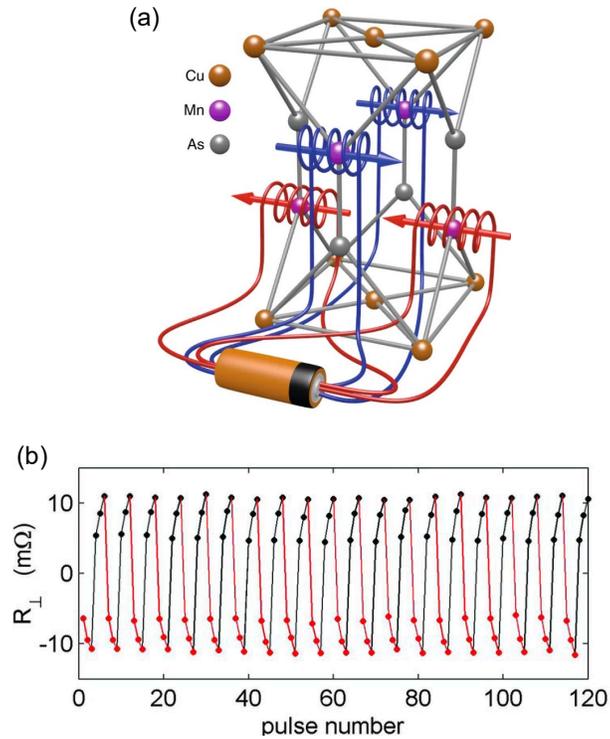

Fig. 11. (a) Illustration of current-induced switching mechanism in the AFM CuMnAs. Staggered magnetic field exists for the two Mn sub-lattices when a current is applied. (b) Measured transverse device resistance as current pulses ($J_c = 4\times10^6$ A cm$^{-2}$) are applied to CuMnAs device. For the black data points, the current was applied along the [100] crystal direction, while it was applied along the [010] crystal axis for the red data points. Reprinted figure with permission from [141]. [copyright statement].

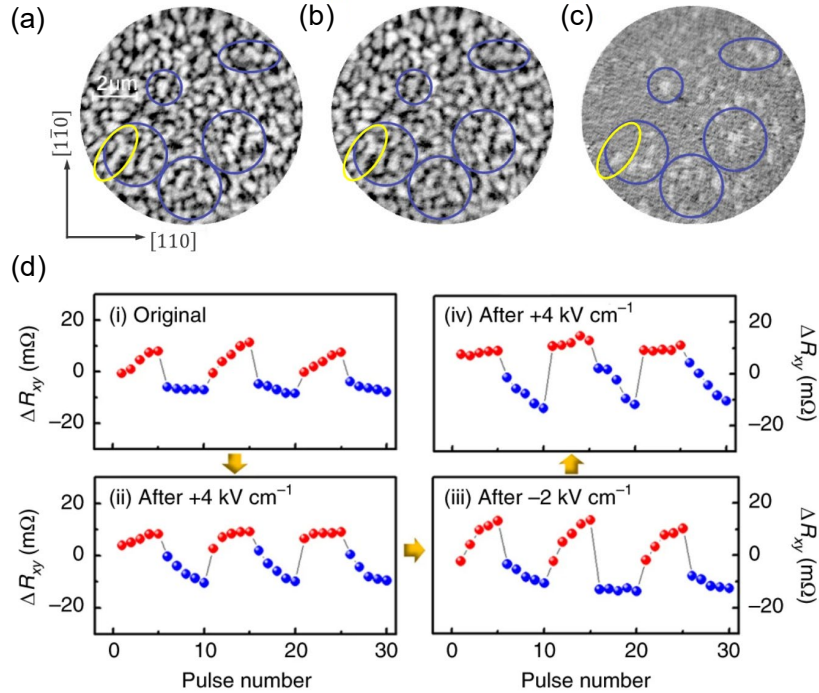

Fig. 12. (a) XMLD-PEEM image of a Mn$_2$Au film before application of current pulses and (b) after applying train of current pulses. (c) Difference of the images in (a) and (b). The blue circle highlights the switched AFM domains. Partially switched domains are highlighted by the yellow circle. Reprinted figure with permission from [149]. [copyright statement]. (d) Current-induced switching behavior of the AFM Mn$_2$Au grown on ferroelectric PMN-PT substrate, measured using transverse resistance. The anisotropy direction was toggled on applying the electric field of opposite polarity to the PMN-PT. Switching with a single pulse is achieved along the easy anisotropy axis while a number of pulses are required to switch along the hard-axis. Reprinted figure with permission from [150]. [copyright statement].

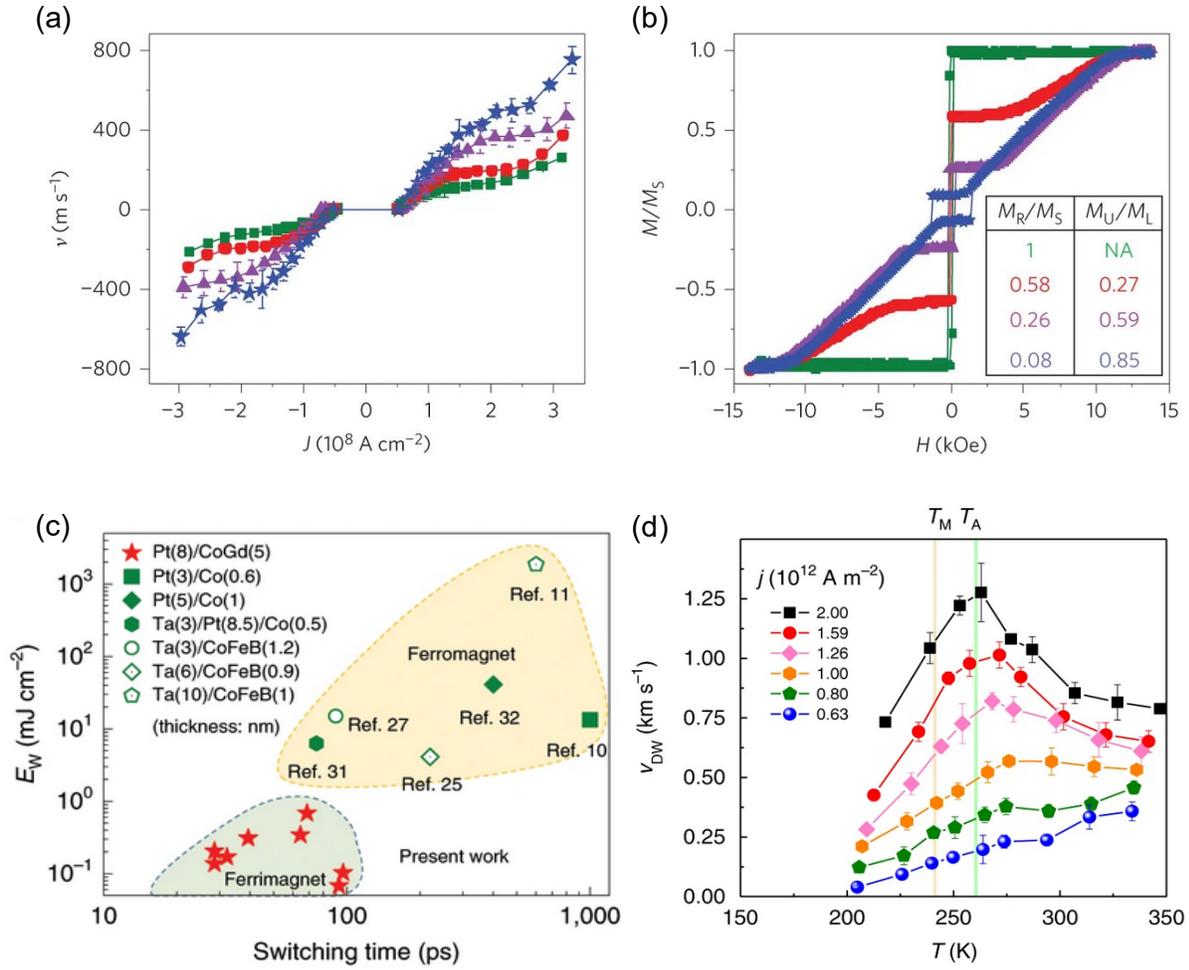

Fig. 13. (a) DW velocity as a function of injected current for Co/Ni/Co based SAF stacks with different compensations. (b) Magnetic characterization of the SAF stack. The green curve is for a spacer (Ru) thickness that results in ferromagnetic coupling (SF – synthetic ferromagnet). Red, violet and blue data points are for SAF with varying degree of compensation achieved by changing the thickness of top layer in the SAF. Reprinted figure with permission from [161]. [copyright statement]. (c) A comparison of switching energy and time between SOT devices based on ferrimagnets (CoGd) and ferromagnets. Please note that the references cited in this figure are the ones listed in the original article. Reprinted figure with permission from [173]. [copyright statement]. (d) Current-induced DW velocity in a CoGd based SOT device as a function of temperature. The highest velocity is obtained at angular momentum compensation temperature. Reprinted figure with permission from [176]. [copyright statement].

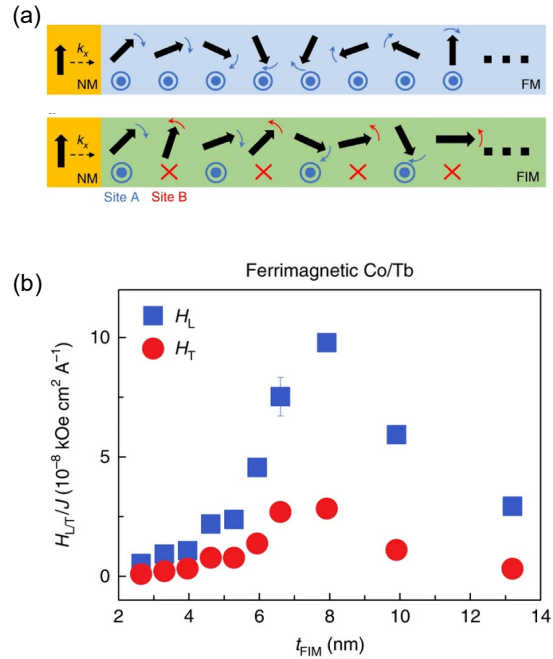

Fig. 14. (a) The precession of injected spin in a FM (top panel) and a FIM (bottom panel). In a FIM, the angular momentum lost by the spin on interacting with first sub-lattice (Site A) is gained back during the second interaction at Site B. (b) Longitudinal and transverse effective fields ($H_L$ and $H_T$) as function of FIM (Co/Tb multilayer) thickness. The long coherence length results in transfer of spin-torque over a larger distance in FIM. Reprinted figure with permission from [178]. [copyright statement].

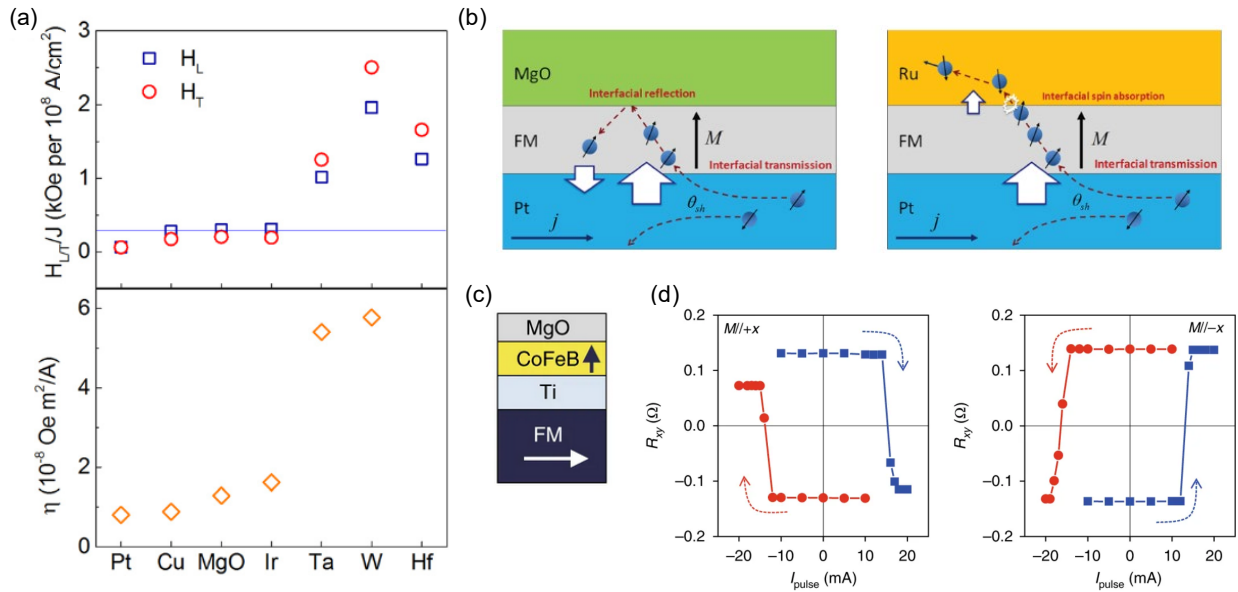

Fig. 15. (a) SOT effective fields and switching efficiency for Pt/Co/[Ni/Co]$_2$ device with different capping layers. Reprinted figure with permission from [187]. [copyright statement]. (b) The spin injected from Pt are reflected back into the FM by an MgO capping while they are absorbed for Ru capping resulting in enhanced SOTs. Reprinted figure with permission from [181]. [copyright statement]. (c) SOT device with an in-plane magnet underneath. (d) Field-free current-induced switching loop for two different magnetization directions of the in-plane FM layer. Reprinted figure with permission from [196]. [copyright statement].

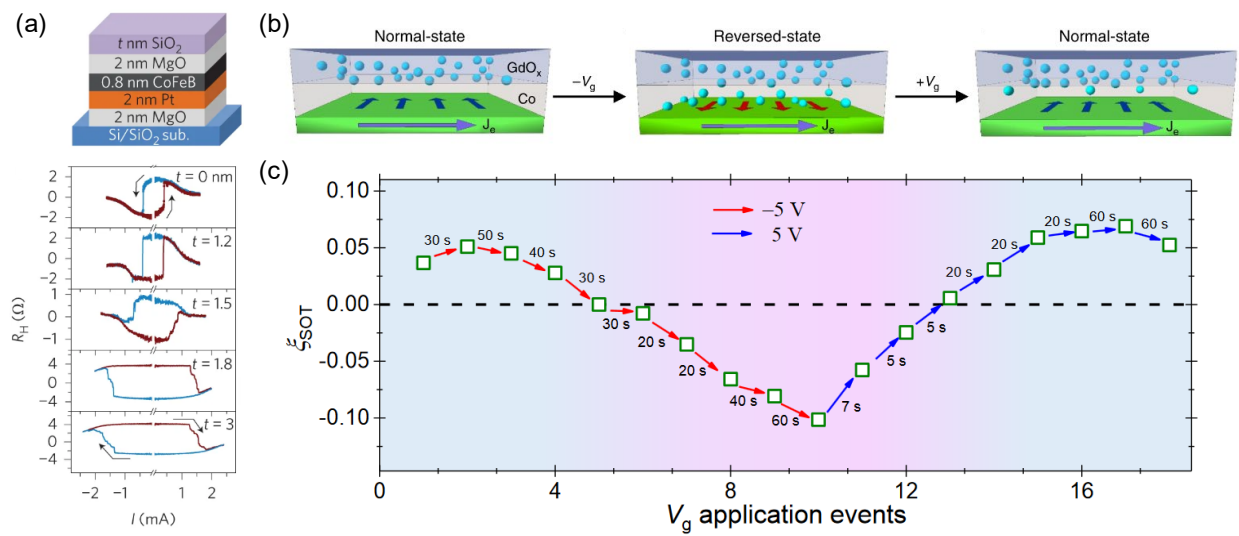

Fig. 16. (a) Device stack and current-induced switching loop for the devices with different $SiO_2$ capping thicknesses. Reprinted figure with permission from [208]. [copyright statement]. (b) Mechanism of electric field induced SOT polarity reversal. The oxygen atoms migrate to the Pt/Co interface (middle panel) on application of negative gate voltage ($V_g$) resulting in reversal of SOTs. (c) Non-volatile programming of the Pt/Co/GdO$_x$ device with different values of effective spin Hall angle, $\xi_{SOT}$. The state of the device is reversible. Reprinted figure with permission from [209]. [copyright statement].

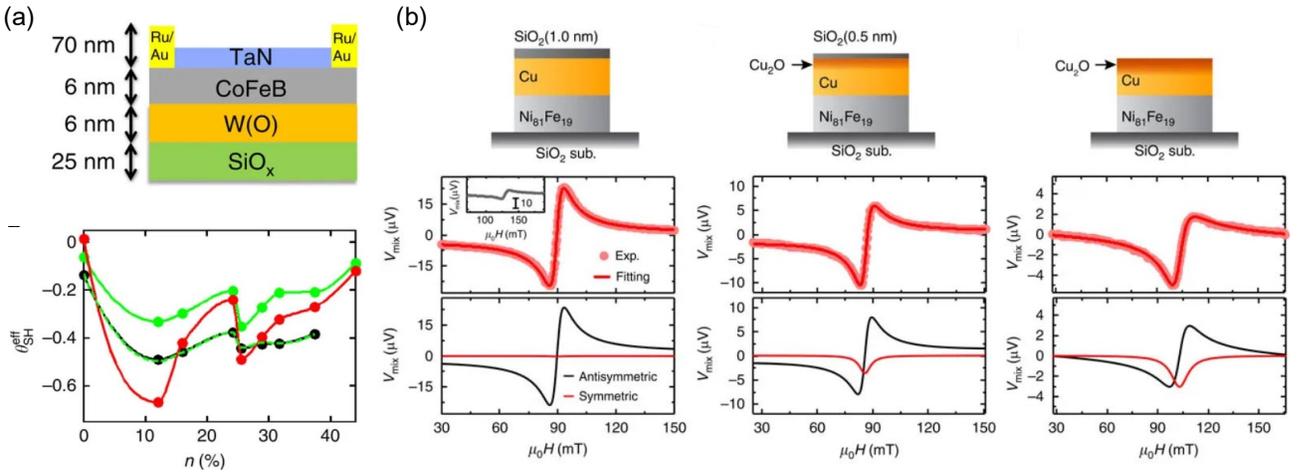

Fig. 17. (a) Oxidized tungsten based SOT device. Effective spin Hall angle ($\theta_{SH}^{eff}$) evaluated using two different ST-FMR analysis techniques as a function of oxygen content in the W layer. The black and red curve are obtained by analyzing the line-width and line-shape, respectively, of the ST-FMR spectra. The line-shape results are corrected by measuring the radio frequency (RF) current using a network analyzer (green curve), followed by further correction of the RF current (dashed-green curve). Reprinted figure with permission from [212]. [copyright statement]. (b) Three different devices with different degrees of surface oxidation in Cu (top panel). The bottom panel shows the ST-FMR spectra for these devices and their fitting with a symmetric and antisymmetric Lorentzian. The symmetric component represents the current-induced torque which increases with increasing Cu surface oxidation for the device. Reprinted figure with permission from [215]. [copyright statement].

## IV. Spin devices and non Von Neumann computing

The present-day computing systems are based on the Von Neumann architecture in which the memory and processing units are separated and information processing is carried out serially. This architecture and computing methodology have fueled the information technology revolution for the past few decades. However, the massive increase in the amount of data with the recent rise of interconnected devices necessitates a new type of computing scheme that can efficiently interpret the data similar to a human brain. To this end, neuromorphic or brain-inspired computing aims at developing devices and circuits that can perform tasks involving learning, training, recognition and cognitive ability. In addition, there is devoted research on alternate computing systems such as Ising machine and quantum computing, which can perform certain optimization tasks at a much faster speed compared to modern computers. In this section, we will discuss recent progress in the spintronic devices and systems that have applications in the above mentioned non Von Neumann computing methodologies.

In the area of spin based neuromorphic computing hardware, a variety of synapse and neuron models have been proposed based on combination of magnetic domain walls, MTJs, and SOTs [219]. The earliest proposal for spintronic synapse shown in Fig. 18(a) has a magnet with a DW acting as a synapse which is connected via a non-magnetic channel to the magnetic neuron [220]. The position of the DW in the synaptic-magnet determines the spin polarization of the current when a voltage is applied on this magnet. The spin polarization is therefore an analog function of the DW position, and thereby represents the synaptic weight. During the write operation a current passed vertically through the synaptic-magnet will carry the weighted information in the form of degree of spin-polarization. This spin polarized current which represents a potential stimulus at the input of neuron can be used to switch an adjacent magnet that acts as a neuron. Many of the these synaptic-magnets can be connected as fan-in to a single neuron which will receive weighted sum of spin currents from these inputs as shown in Fig. 18(b).

In a later proposal, Sengupta et al. proposed a synaptic design which has a DW integrated with a MTJ as shown in Fig. 18(c) [221]. In this structure, the weight is written by SOTs induced DW motion. The DW moves transverse to the channel length when a current is passed from terminal C to D as illustrated in the figure. The position of DW in the free layer of the MTJ determines its conductance which can be inferred as the synaptic weight. The parallel alignment of the free layer with respect to the reference layer results in a high conductance or maximum weight. On the other hand, an anti-parallel alignment between two layers is a state of minimum weight due to a low conductance of the MTJ. The rest of the conductance states that lie in between these two states are determined by the DW position in the free layer. A similar device structure but without the extended free layer and HM channel (see Fig. 18(d)) was also proposed to perform identical synaptic functions [222]. In this design, the current is applied along the length of the MTJ and the DW moves along the current direction. The proposed device can emulate a neural transfer function as well when connected with a reference MTJ and a transistor as shown in Fig. 18(e). The reference MTJ in series with the synaptic or DW MTJ acts as a voltage divider between the supplied voltage and the ground. The position of DW in the synaptic MTJ determines the divided voltage at the input terminal of the transistor. The resulting transfer function of the complete device is in form of a sigmoid as shown in the inset of Fig. 18(e). It should be noted that a sigmoidal transfer function forms a building block of artificial neural network (ANN). Similar to the proposals of spin logic in Fig. 3(a), the above proposals are also based on ideal behaviors of the DW, the magnet and the spin interconnects. The superposition of spin currents from different fan-ins as shown in Fig. 18(b) has not been demonstrated yet. Moreover, precise optimization of interconnect lengths and materials is also required to avoid spin decay. For DW based synapse, the nano-dimension of the future devices will result in very few analog states. For further details on a potential spin based hardware solution for neuro-computing we refer to the reader a focused review [219].

On the experimental side of spintronic neuromorphic hardware, Lequeux et al. have demonstrated a DW based synapse which has a large number of intermediate resistance states [223]. This spintronic synapse consists of a MTJ with a single DW in the free layer. The DW can move back and forth with the aid of STT applied using a positive and negative current pulse. The pinning of the DW results in an intermediate domain configuration, hence in-between resistance states (Fig. 19(a)). A problem with this device is a reliable control of the DW. The motion of DW is not predictable due to the inhomogeneity of the material and thermal effects. While a linear weight tunablity is desirable for synapse in ANNs, the DW synapse has a non-linear weight programming due to the arbitrary distribution of the pinning sites. A series of MTJs with moving DW under them have been used to implement precise and linear weight generator [224]. The schematic of this device and the corresponding scanning electron microscope image is shown in Fig. 19(b). The variation of the MTJ device areas produces a non-linear activation function as shown in Fig. 19(c). Borders et al. have used a AFM-FM (PtMn-[Co/Ni]) bilayer SOT device which presents analog switching states like a biological synapse (Fig. 19(d)) [225]. The analog switching behavior is due to pinning of domains with AFM grains. In this work, the authors have demonstrated associative memory operation using 36 of these SOT devices integrated with a field programmable gate array.

Recently, magnetic synapse has been demonstrated with a Pt/Co/GdO$_x$ SOT device [226]. Its working principle involves electric field induced reversible oxidation and reduction of the Co layer, thereby modulating its magnetization. The magnetization of the Co layer represents the synaptic weight measured using the anomalous Hall resistance ($R_{AHE}$) as shown in Fig. 19(e). Synaptic functionalities like potentiation, depression, spin-rate and -time dependent plasticity were demonstrated on this device. Fig. 19(e) shows the effect of different stimulating pulse frequency on the synaptic weight. Similar to a biological synapse, frequent stimulations result in a larger change in the synaptic weight when compared to sporadic stimulations. Since the device is based on oxygen migration, the speed of synaptic update is quite slow. The read-out of the magnetization using the anomalous Hall resistance is also non-ideal for supporting a large number of fan-outs which are present in a typical neural network. Magnetic skyrmions have been also used to mimic the potentiation and depression function of biological synapse [227]. The number of skyrmions in the magnetic channel determine the weight of the synapse which can be measured using the anomalous Hall resistance. Creation or annihilation of a single skyrmion in a ferrimagnetic film using a current pulse results in a quantum jump of the synaptic weight similar to the potentiation or depression of biological synapse. Since the number of skyrmions in the channel can be effectively controlled using current pulses, corresponding weight update is approximately linear as shown in Fig. 19(f). However, the skyrmion readout signal is generally very small, which poses a challenge in the readout operation and dynamic range of these synapses.

While the synaptic functionalities using spintronic elements have been implemented by a few groups, a hardware realization of neuron has been rather challenging and limited. Few experiments have used MTJs to demonstrate a stochastic spiking neural function. A threshold current to drive the MTJ in an unstable state that results in stochastic current spikes was used in one of the scheme [228]. In a recent experiment, VCMA was used to enable a stochastic switching behavior of the free layer of the MTJ as shown in Fig. 20(a) [229]. In this work, the perpendicular anisotropy of the free layer was modified by an applied bias voltage to the MTJ. The bias voltage results in lowering the energy barrier leading to a stochastic switching of the free layer. The switching probability has therefore a sigmoidal relation with the applied bias field.

While a stochastic device with sigmoidal transfer function is suited for ANN, the leaky integrate-and-fire neuron is necessary to implement a spiking neural network. Thermally assisted current-induced SOT switching was exploited to enable this integrate and fire function [230]. Current pulses arriving at a high frequency integrate the temperature-current budget above the switching threshold of the SOT due to minimal leakage or less heat dissipation. This results in magnetization switching or neural firing (probability of switching, $P_{sw} = 1$). For current pulses arriving far apart, the heat generated by the first pulse is dissipated by the time when the second pulse arrives, therefore the SOT device does not switch ($P_{sw} = 0$). For, pulses of intermediate frequency, $0 < P_{sw} < 1$. Figure 20(b) shows the distribution of switching probability as a function of pulse frequency for the given SOT device. This behavior is similar to a biological neuron in which incoming potential spikes to a neuron integrate in a leaky fashion resulting in neural firing when the membrane potential exceeds a pre-determined threshold. However, it should be noted that while the above SOT neuron can integrate and fire, it does not automatically reset like a biological neuron. This functionality is yet to be achieved using a spin device.

Apart from the research on spintronics counterpart for biological synapse and neuron, there are some interesting works on system level implementation of spin devices for recognition and optimization tasks. Torrejon et al. have used a spin-torque oscillator (STO) which converts an input dc current into voltage oscillation through TMR, for spoken vowel recognition [231]. A spin-torque oscillator is basically a MTJ in which the magnetization oscillates in a self-sustaining fashion around the effective magnetic field when a balance is achieved between the magnetic damping and applied current-induced torque. The oscillation of the magnetization in effect results in an oscillating TMR signal. The frequency of these oscillations is a function of the effective magnetic field experienced by the magnet and its gyromagnetic ratio. Therefore, a STO is characterized by its oscillation frequency which can be tuned by varying the external magnetic field. A STO combines non-linearity and memory in a single device [232]. The non-linear behavior is between the applied current and the amplitude of generated voltage, while the memory function is achieved due to the dependence of output on the past input currents. After pre-processing, the input speech signal was applied in the form of current to the STO, while the output was recorded in form varying voltage signals. A comparison of the recognition performance with and without an oscillator in Fig. 21 shows clearly the improved performance of the STO based system.

In a later work, four of these oscillators were electrically coupled for a vowel recognition task [233]. The electrical coupling was achieved by physically connecting these oscillators with wires. These oscillators were synchronized with two external microwaves as shown in Fig. 22(a). Each oscillator synchronizes itself with an external microwave frequency in a different range as shown in the right panel of Fig. 22(a). The range of synchronization can be tuned by varying the applied bias current to the individual oscillator. The spoken vowels were coded as a function of two external input microwave frequencies ($f_A$, $f_B$). The frequency distribution map of each vowel for different speakers overlapped on the oscillator synchronization map is shown in the left panel of Fig. 22(b). The color in the synchronization map represents the oscillator or oscillators synchronized with the two microwaves as shown in the legend bar to the right of the figure. For example, (1A) represents the 1st oscillator synchronized with microwave source A, and (2A, 4B) represents 2nd and 4th oscillator synchronized with microwave source A and B, respectively. Since the goal of this work was to recognize vowel independent of speaker, the learning involved adjusting the bias current through the oscillators so that the points for each vowel are contained in single synchronization region. The synchronization map after different training steps are shown in the middle and right panels of Fig. 22(b). Comparing the first and last panel of Fig. 22(b), it can be seen that after sufficient training, the spoken vowels which were initially distributed in more than one region on the synchronization map, eventually reside in approximately a single synchronization region. A reasonable recognition rate of 89% was achieved after around 50 training steps. The disadvantage of using oscillators for neuromorphic computing is their limited scaling potential. In addition, the frequency spectrum of spintronic oscillators is not very sharp and has a large full width at half maximum which hinders their operational reliability.

Very recently, the stochastic behavior of a thermally unstable MTJ has been utilized to develop a probabilistic-bit (p-bits) for integer factorization [234]. A p-bit is an entity that fluctuates between two binary states with a probability that can be controlled by an input [235,236]. The relation between the input $I(t)$ and output $m(t)$ of a p-bit is given by

$$m(t) = \text{sgn}[\tanh(I(t)) - \text{rand}(-1, 1)], \tag{1}$$

where rand(–1, 1) is a random number uniformly distributed between –1 and 1 [235]. The stochastic nature of the p-bit finds applications in probabilistic computing. The MTJ for a spintronic p-bit is designed by optimizing its volume and free layer thickness so that the energy barrier between its two bistable states is low enough to be surpassed by ambient thermal energy. This results in the MTJ switching stochastically. Initially, current controlled MTJ p-bits have been proposed [235,237] as shown in Fig. 23(a). These three-terminal p-bits were driven by SHE torques induced by input currents flowing in the heavy metal below the MTJ. The output can be read by passing a small read current through the MTJ which can be fed into a buffer stage. The CMOS buffer stage helps in providing a gain and isolation at the output. For the magnitude of input current greater than the threshold value the MTJ free layer pins in one of the stable states resulting in an output of +1 or –1. However, for values of input current in between the MTJ behaves stochastically and the instantaneous value of the output can fluctuate between –1 and 1 as determined by Eq. (1).

In a voltage-controlled scheme, the stochastic MTJ is connected to a n-type transistor (NMOS) as shown in Fig. 23(b) [234]. The transistor also has a resistance ($R_{source}$) at its source terminal which limits the currents through the MTJ to a value at which its switching probability is 0.5 [234]. The output of the transistor is connected to a comparator which determines the final output ($V_{OUT}$) of the p-bit. While the instantaneous $V_{OUT}$ of the p-bit will be either of the rail-to-rail values, the time averaged output (over 700 ms and 2000 sampling points) is a sigmoidal with respect to $V_{IN}$ as shown in Fig. 23(c). For performing factorization, these p-bits are interconnected (not physically) such that the input of each p-bit is a function of output of all the other bits. In general, the input $I_i$ for the $i^{th}$ bit is determined by $-\delta E(m_1, m_2 ..)/\delta m_i$. Here $m_i$ is the output of $i^{th}$ bit and $E(m_1,m_2..)$ is the energy cost function. The inputs are calculated analytically from the outputs and the cost function ($E$) using a microcontroller. It should be noted that depending on the optimization problem, the associated cost function will be different. The p-bits frequently end up in configurations that minimizes the cost function, $E$.

Figure 23(d) shows a factorization result of number 35 obtained using 4 p-bits. The two factors in binary form are represented by ($p_1$, $p_2$,1) and ($q_1$, $q_2$,1), where $p_i$ and $q_i$ are the four p-bits. The top panel of the figure shows an equal distribution between different states of the p-bits when they are uncorrelated, in this case, their inputs function being zero. When the p-bits are connected, their probability of settling in a configuration is highest for the values representing the numbers of (5, 7) and (7, 5), i.e. factors of 35, as shown in the bottom panel of Fig. 23(d).

Occasionally the p-bits end up in state representing the number (5,5), (7, 7), etc., although with a very less probability. In this work, spin based probabilistic computing compared to CMOS-based alternatives was suggested to offer an energy benefit of 10 times and an area advantage of 300 times. However, whether the proposed scheme can be scaled for a real application is questionable (e.g. the most common form of 256-bit encryption corresponds to 78 digits). It should also be noted that the network weights are still implemented in a microcontroller in this work. Going forward these should also be implemented in hardware using memristors or capacitive network.

Other application of p-bit enabled stochastic circuits include machine learning inspired applications such as Bayesian inference and accelerating learning algorithms. Quantum inspired applications such as inverted Boolean logic (e.g. finding input to a logic gates for a given output) and optimization problem like travelling sales man problem can be also solved using a network of p-bits [238]. Current-controlled p-bit can be used to implement binary activation function in a binary neural network (BNN) in combination with memristors that implement weights [239]. A more detailed review on these applications can be found in [238].

There are many other alternative computing methodologies and application specific hardware that can be enabled in an energy- and area-efficient way with spin devices. BNNs which use binary values of weights and activations, achieve the same degree of accuracy compared to the normal neural networks while being more resource efficient in terms of storage, speed and power. During inference the BNNs use the XNOR operation, which can be implemented using a single SOT device. The two inputs of the write driver of a SOT cell can serve as the inputs to the XNOR gate [240]. Non-volatile memories are also used for in-memory computations, which help in saving a large amount of power and time that is spent due to the separation of memory and computing units in a von Neumann architecture. The in-memory computing capability has been proposed to implement a two-bit AND gate used for performing the bit convolutions in BNNs [240,241]. In this scheme, the row decoder is modified to turn on two read-word lines simultaneously. The sense current of the selected SOT bits flowing through the bit line determines the values stored in them. By appropriately setting the reference voltage in the sense amplifier circuit, logic AND can be implemented. Instead of using two bits, a single STT cell-based scheme was proposed for performing in memory computation by fabricating two MTJs on top of one other [242]. Apart from the use in BNNs, the in-memory computing architecture using spin devices can be leveraged for bioinformatics (e.g. DNA read alignment) [243,244] and graph processing applications [245,246].

The results discussed in this section demonstrate the viability of spintronics hardware for non-von Neumann computing systems. At present the ANNs are implemented mostly in the normal computers. Since the weights of the network is stored in memory and the computing is performed in the processor, the flow of information between these two components is both a speed and power bottleneck. Hardware implementations of ANNs have involved accelerators for matrix multiplication e.g. GPUs and in few cases using transistors circuits to implement weights. However, these efforts are not power and area efficient. The biggest stride in the use of alternative solid-state devices to build components of ANNs is using memristive technology [247,248]. A single memristor incorporates most of the functionality of a synapse that is used in both ANN and biological neural networks. The most important feature of these functions is weight programmability in which the weights are represented by the conductance of the memristor. The simple device structure of a memristor i.e. an insulator between two metals, also enables 2D and 3D cross-point architecture that is very efficient in performing matrix multiplication (between inputs and weights) [249,250]. However, the operation principal of memristive devices which is mostly based on stochastic movement of ions exposes them to cycle to cycle variability. In addition, the memristor physics is still not well understood and modeled, therefore there is device to device variability that makes the scaling of memristive architecture challenging. While the effort towards spin devices for neuromorphic computing is fairly recent and in an early stage, the well-established physics and fabrication process of spin devices gives a larger controllability when used in neuromorphic applications. The weight in a spin device is expressed in the form of its magnetization which is non-volatile and can assume analog values similar to memristive conductance. Since the process of weight modulation in different types of spin devices does not involve physical motion of ions (for most cases), spin devices have a high durability and endurance compared to the memristor counterparts. Future efforts involving spin devices in ANNs should focus on building memristor-like cross-point architectures to demonstrate its scalability.

Going forward it is also necessary to integrate spin based synapse and neurons for a complete neural architecture. In order to achieve general purpose computation, we may need to mimic biological systems such as a human brain, in which one neuron has 10,000 synaptic connections. Obviously, this requirement is beyond our capability using modern fabrication techniques. If we aim to solve a specific problem, however, a typical cross-point architecture can fulfill the job. For the time being, research activities will be focused on developing a particular recognition and optimization hardware such as the ones discussed in ref. [231,233,234] accelerating the field towards real applications. The learning part of the recognition task which is still performed on conventional computer should also be enabled on spin architecture in order to realize a full spin based non Von Neumann computing system.

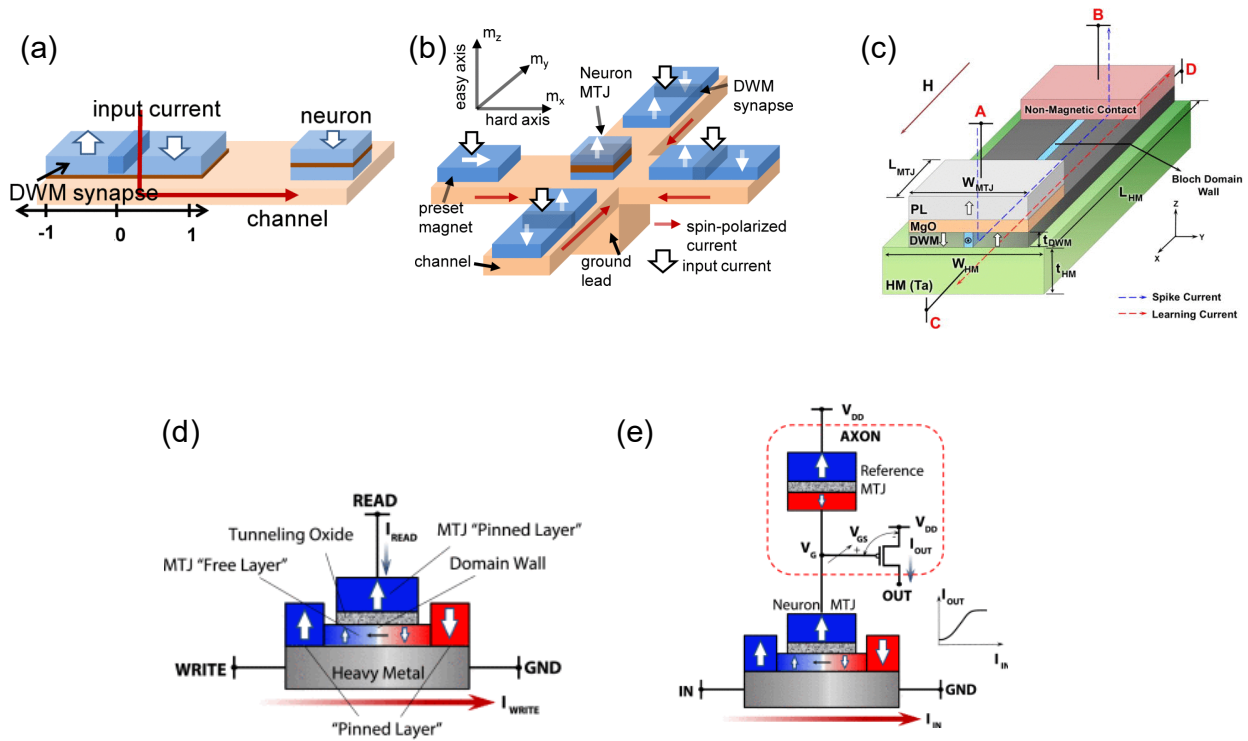

Fig. 18. (a) Illustration of a DW based spintronic synapse. (b) Spin-neural network based on DW based spin synapse. Reprinted figure with permission from [220]. [copyright statement]. (c) A DW based spin synapse in which the synaptic state is written by passing current through the channel and read-out is performed using a MTJ. Reprinted figure with permission from [221]. [copyright statement]. (d) Current based writing and MTJ based reading for a spin synapse (left panel). A connected network of these synapses (right panel) to perform the weighted summation of inputs. (e) A DW-MTJ based spin neuron to enable sigmoid functionality. Reprinted figure with permission from [222].

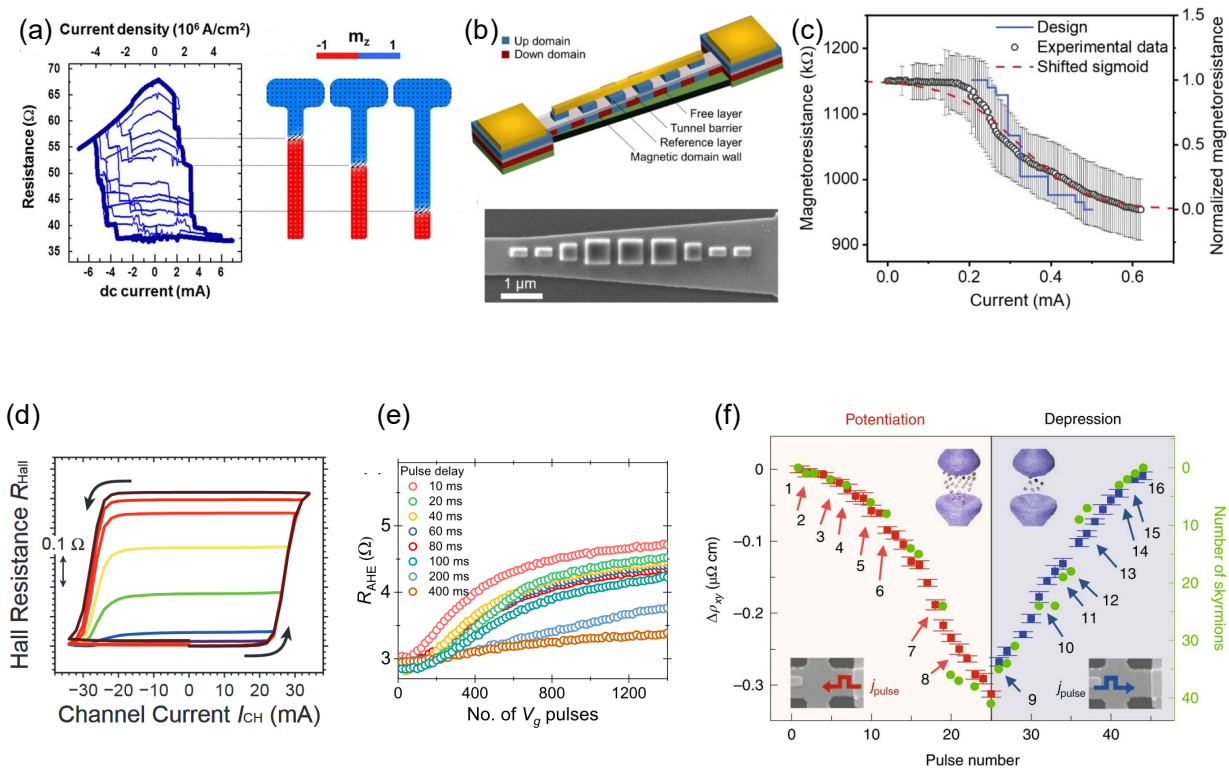

Fig. 19. (a) Various resistance states of a DW synapse programmed using STT. The right panel illustrates the domain configuration for the different device states. Reprinted figure with permission from [223]. [copyright statement]. (b) Schematic of spintronic synapse which has series of MTJs for programming analog weights. Bottom panel shows the SEM image. (c) The non-linear conductance behavior obtained with the series-MTJ based synapse containing MTJs with varying dimensions. Reprinted figure with permission from [224]. [copyright statement]. (d) A PtMn based SOT device with programmable synaptic weights. Reprinted figure with permission from [225]. [copyright statement]. (e) The programmed synaptic weight for a Pt/Co/GdO$_x$ based spintronic synapse depending on voltage pulse frequency, similar to a biological synapse. Reprinted figure with permission from [226]. [copyright statement]. (f) Potentiation and depression of skyrmion based synapse with current pulses. The synaptic weight is measured using anomalous Hall resistance. Reprinted figure with permission from [227]. [copyright statement].

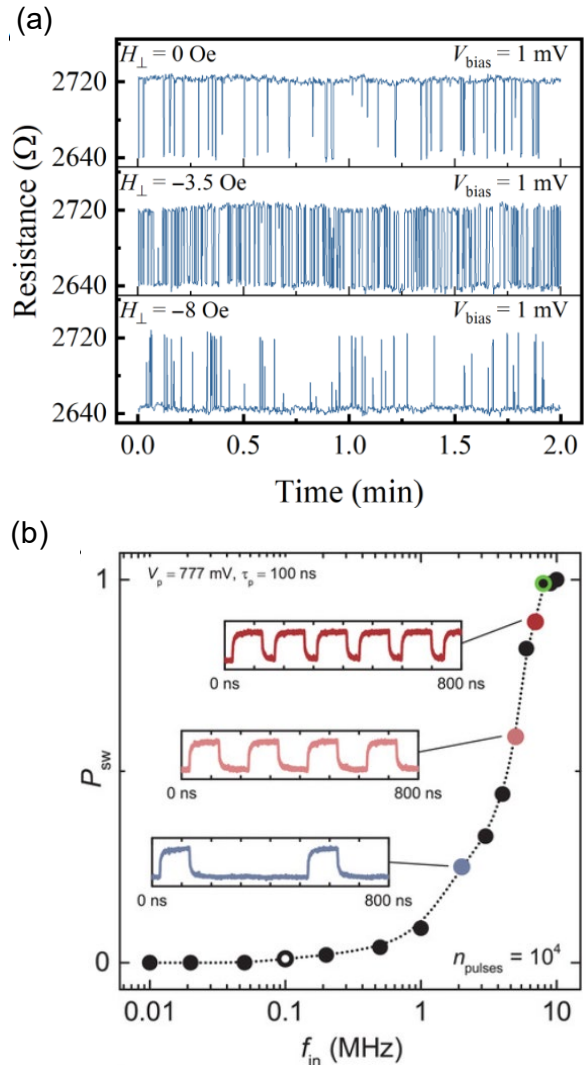

Fig. 20. (a) Voltage-induced stochastic behavior of a MTJ device achieved by controlling the perpendicular anisotropy. The behavior resembles the firing of biological neuron. Reprinted figure with permission from [229]. [copyright statement]. (b) Switching probability as a function of pulse frequency for a PtMn based SOT device. The device performs leaked integration which is an essential neuron property. Reprinted figure with permission from [230]. [copyright statement].

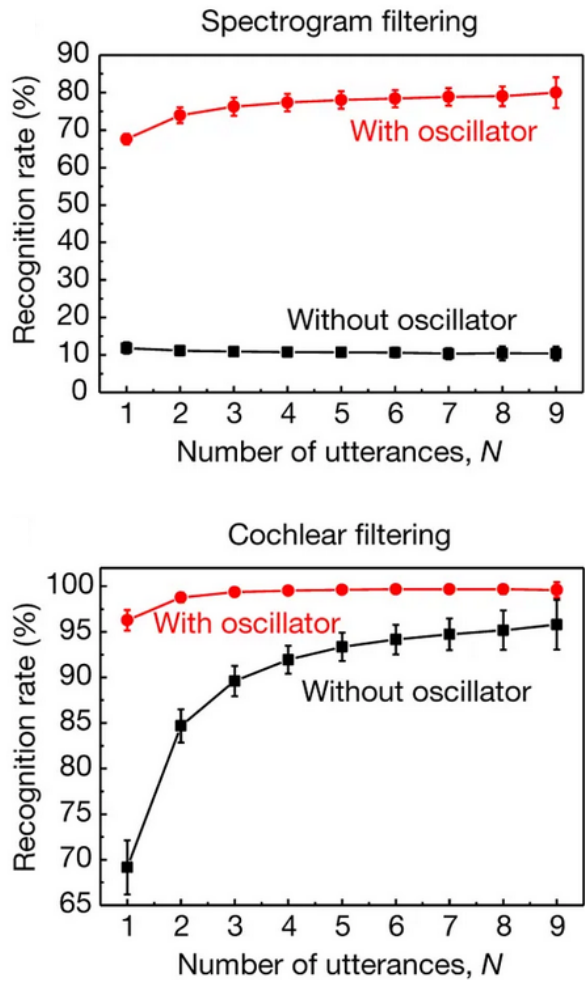

Fig. 21. Recognition rate without and with an oscillator as function of number of utterances used for training. Two different filtering techniques were used for processing the input audio file. The recognition without an oscillator was performed on computer. Reprinted figure with permission from [231]. [copyright statement].

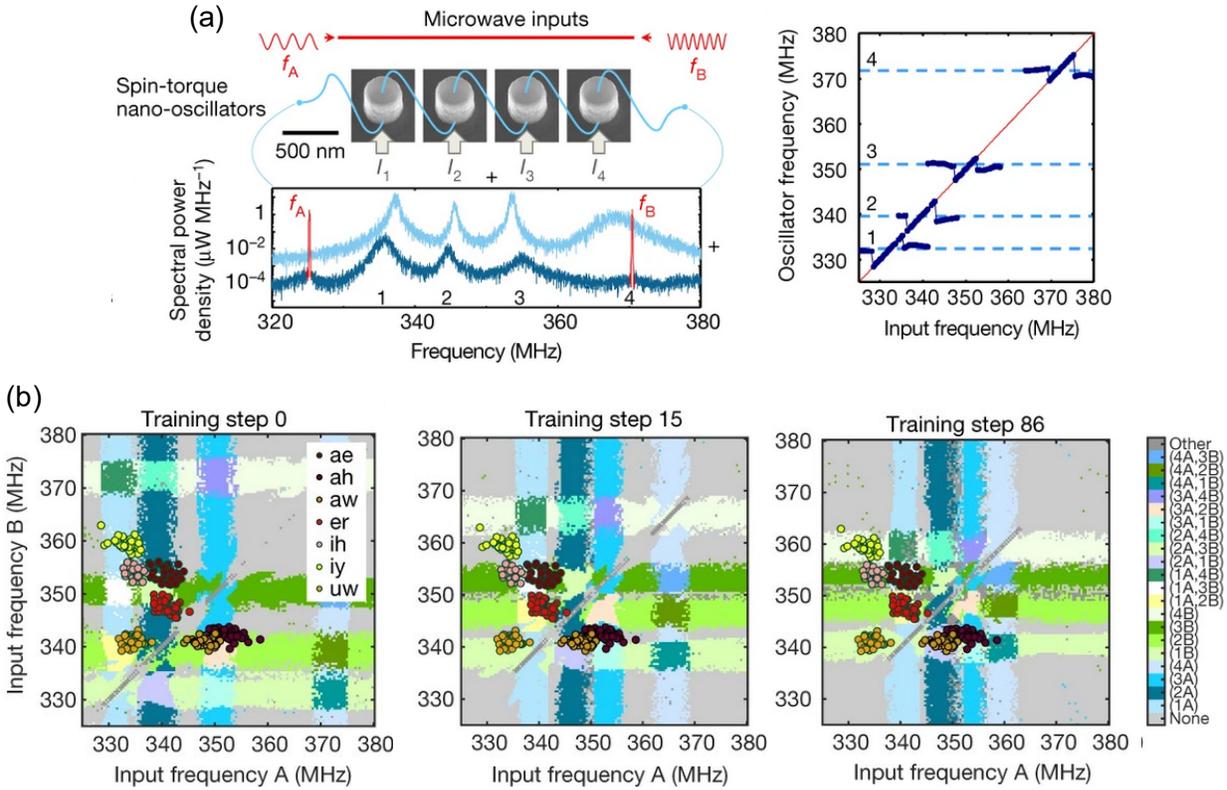

Fig. 22. (a) Experimental setup of four electrically connected and coupled spin oscillators. Two external microwaves with frequencies $f_A$ and $f_B$ are applied to the coupled system. The microwave outputs from the oscillators without and with the external microwaves are shown by the light blue and dark blue graphs, respectively. The right panel shows the synchronization of individual oscillator with external microwave. (b) The color code background in the figures shows the synchronization map of the oscillator as a function of $f_A$ and $f_B$. The color coding represents the oscillator or oscillators which are synchronized (see legend). The data points are frequency coded representation of vowels spoken by different speakers. The evolution of the synchronization map after 15 and 86 training steps is shown in the middle and right panels, respectively. Reprinted figures with permission from [233]. [copyright statement].

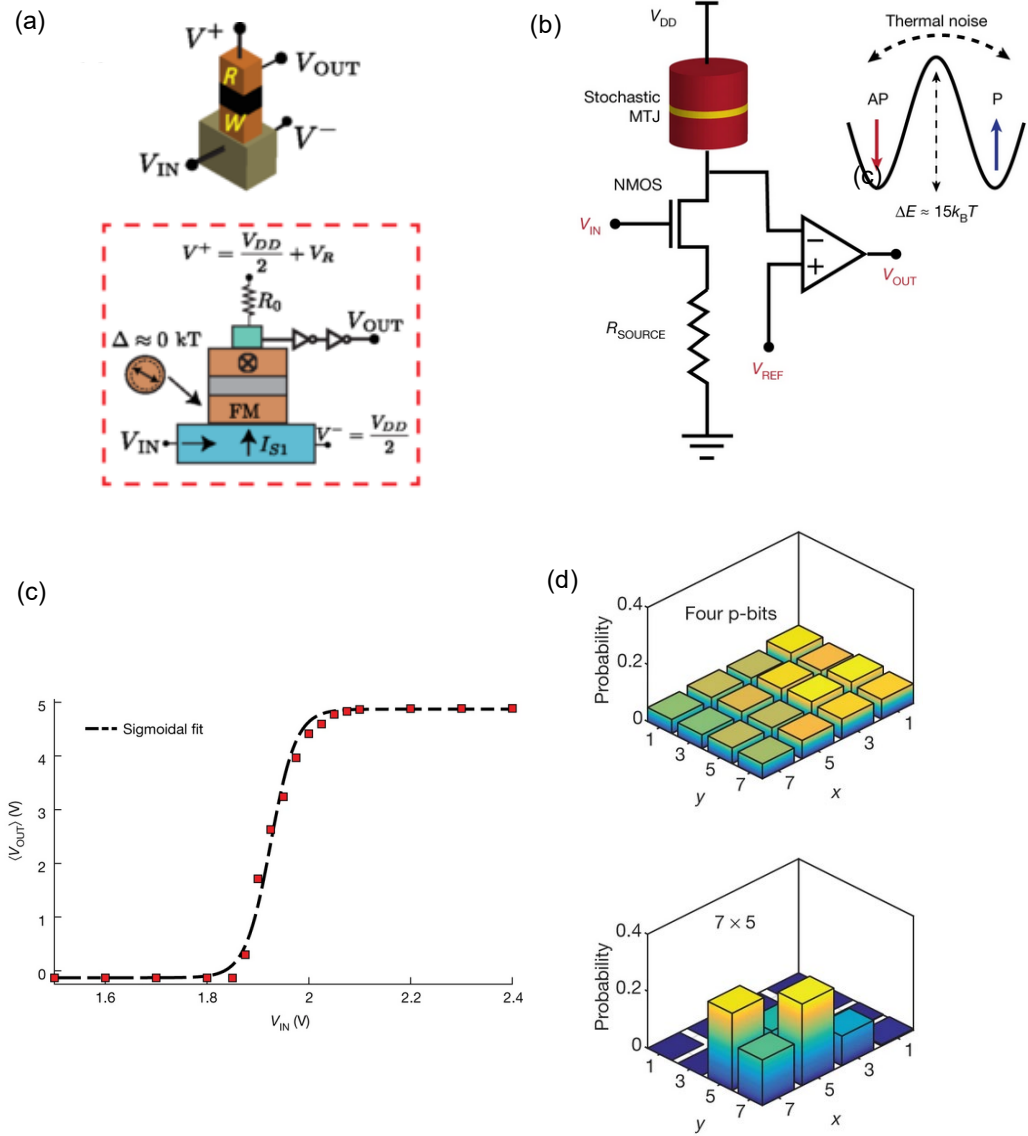

Fig. 23. (a) A current controlled probabilistic or p-bit. The MTJ has a low thermal barrier at room temperature which allows it to randomly switch between the two stable states. The stimulus is applied using the current induced SHE. Reprinted figure with permission from [235]. [copyright statement]. (b) A voltage controlled p-bit based on a stochastic MTJ, a transistor and a comparator. (c) The time averaged output and input voltage of the p-bit follows a sigmoid relationship. (d) The uncorrelated and correlated state of the 4 p-bit system is shown in the top and bottom panel, respectively. The correlate states are used to factorize the number 35. Reprinted figure with permission from [234]. [copyright statement].

## V. Spintronics for flexible electronics

A huge segment of the future generation of consumer electronics such as wearables, medical implants, displays, etc. depends on the fabrication of electronics components on flexible substrates. A key requirement is that the device performance should be comparable to that when they are fabricated on conventional rigid substrates. In the field of spintronics, the deposition of exchange-biased magnetic layers has been shown on free-standing organic films (e.g. mylar, Kapton, ultem, etc.) a few decades back [251]. Later [Co/Cu] based GMR multilayers deposited on plastic substrates with a photoresist buffer were shown to have two times larger GMR compared to the films deposited on bare silicon substrate [252]. The GMR value of these multilayers was unaffected by tensile deformation of up to 4.5% when grown on elastic poly (dimethylsiloxane) (PDMS) membranes [253]. These GMR layers were also made into printable ink by dissolving the multilayers deposited on photoresist coated silicon films in acetone and subsequently mixing the dissolved flakes in a binder solution [254]. Ota et al. have recently shown that GMR devices can even be used for sensing the direction of strain [255].

Since the MTJs form the backbone of modern spintronic applications, integrating them on flexible substrates has been a topic of active interest [256-259]. Co/$Al_2O_3$/NiFe MTJs have been fabricated on Kapton substrates which have a robust TMR with stress/bending as shown in Fig. 24(a) [258]. However, it has been shown that the TMR can in fact be engineered with strain for in-plane MTJs with an MgO barrier [256,257,260]. In a series of measurements on MTJs fabricated on silicon substrates it was shown that increasing strain results in a two-fold increase in the TMR [257]. The strain on the sample was applied through a clamp and screw setup and the obtained results are shown in Fig. 24(b,c). It was elucidated that while the parallel resistance of the MTJ channel remains same under strain, the antiparallel resistance increases, resulting in a larger TMR.

The TMR remained small and unperturbed by strain for a non-annealed sample, thereby establishing the sensitivity of quantum tunneling through an epitaxial MgO barrier to strain. In a later work, it has been shown that the MTJ stack when fabricated on a flexible polyethylene terephthalate (PET) substrate exhibits stable and reliable TMR values. In fact, the TMR of MTJ on the PET was 50 % higher when compared to Si substrates [256]. The MTJs were fabricated by transfer print process. In this process illustrated in Fig. 25, the MTJs were first fabricated on Si substrate, and the substrate was etched using dry etching methods. The suspended MTJ stack was then transferred on the PET, glass, Al foil, PDMS and nitrile glove (Fig. 26(a)). Figure 26(b) shows that TMR of the device is stable even after application of various degree of stress over time.

Apart from the GMR and MTJ devices, other important spintronic phenomena and devices have been reliably demonstrated on flexible substrates. Vemulkar et al. have shown successful fabrication of antiferromagnetic nanowires down to 210 nm width on flexible substrates [261]. The switching characteristic of both the film and the fabricated device was as robust as the one grown on Si. Wang et al. showed the voltage control of magnetic anisotropy through ionic liquid gating on Pt/Fe/Pt/Ta film on polyimide flexible substrates [262]. Similarly, successful fabrication of Pt/Co SOT devices has been demonstrated on flexible plastic substrates [263]. The SOT devices on flexible substrates show a stable current-induced switching characteristic under both compressive and tensile bend conditions as shown in Fig. 27.

Strain engineering has been extensive applied to modern p-type field effect transistors in order to improve the mobility and to optoelectronic devices to modify the effective hole mass, both of which improved the device performance significantly. In the future, flexible spintronics research should aim towards development of complex spintronic circuits on flexible wafers and integration of silicon components alongside the spintronics counterpart on these flexible wafers. More importantly, active strain engineering to enhance the device performance, detecting the amount of strain, and even harvesting the energy could be envisioned using flexible spintronics devices. A more cross-disciplinary approach which involves adopting learnings from efforts in other flexible electronics counterparts is essential for a faster implementation.

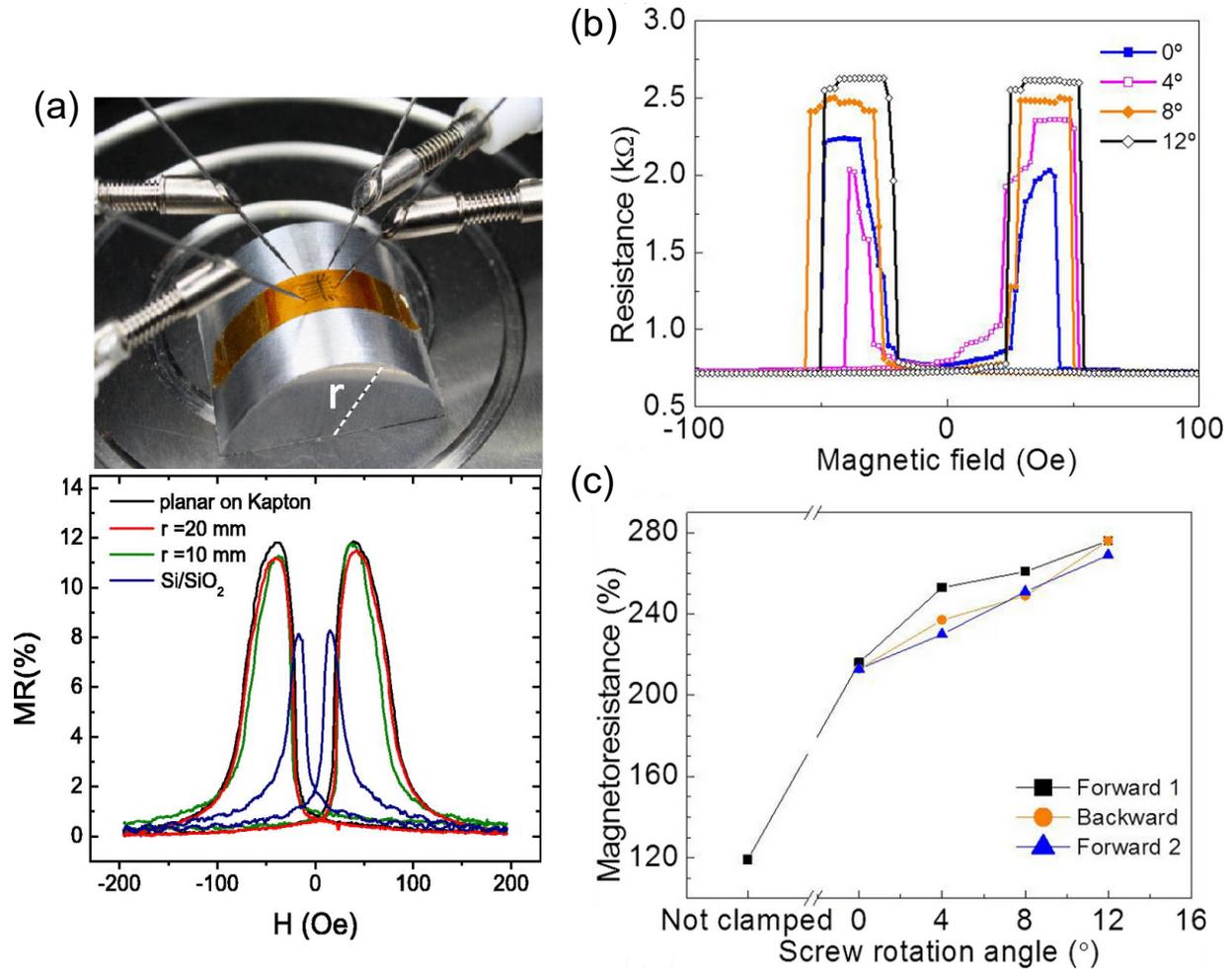

Fig. 24. (a) Measurement setup and the measured TMR for $Al_2O_3$ based MTJ on Kapton. Reprinted figure with permission from [258]. [copyright statement]. (b) TMR on MgO based MTJ deposited on $Si/SiO_2$ substrate. The substrate was clamped and bent using a screw underneath. Legend in the figure represents the amount by which screw was rotated to apply strain on the sample. (c) TMR value as a function of screw rotation or the amount of strain on sample. Reprinted figure with permission from [257] . [copyright statement].

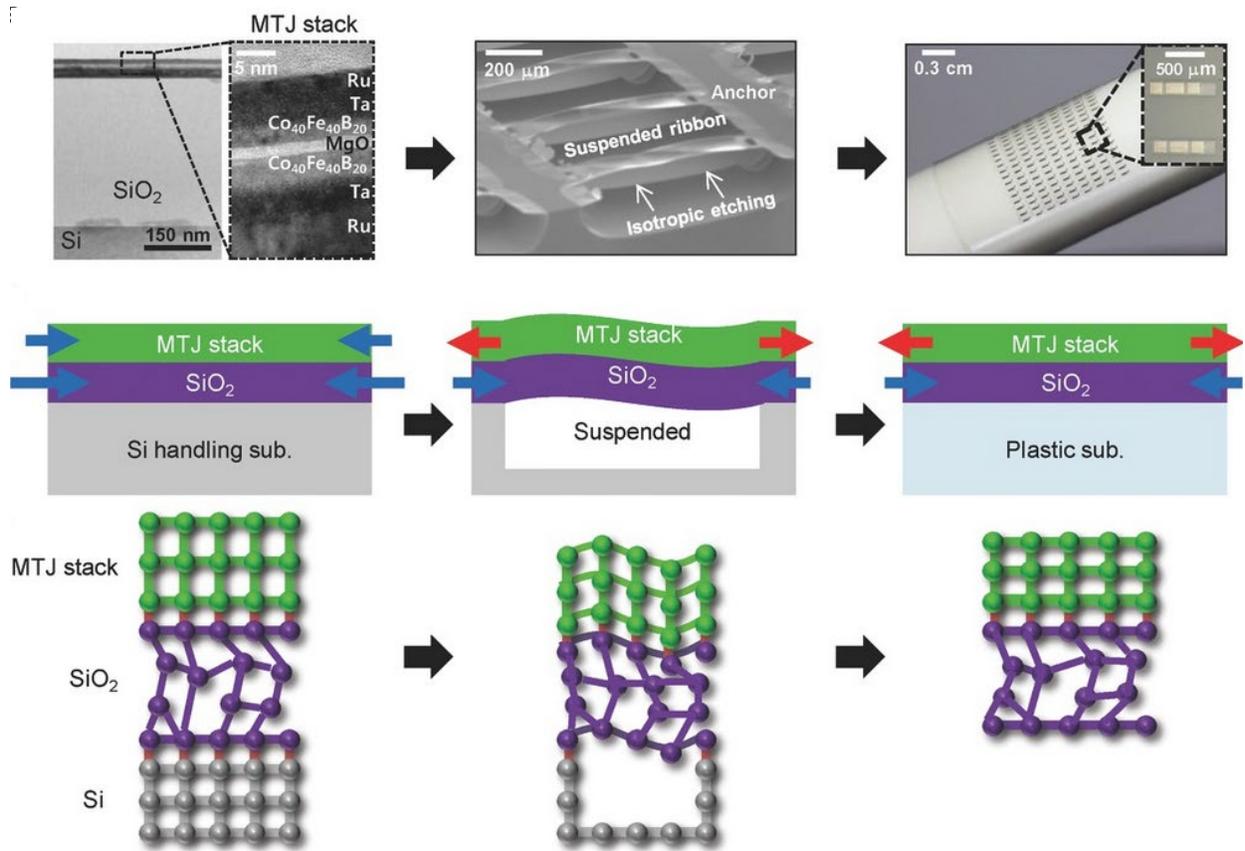

Fig. 25. Schematic of various steps of the transfer print process for fabricating MTJ on flexible substrate (bottom two panels). The top panel shows the actual device during these steps. Reprinted figure with permission from [256]. [copyright statement].

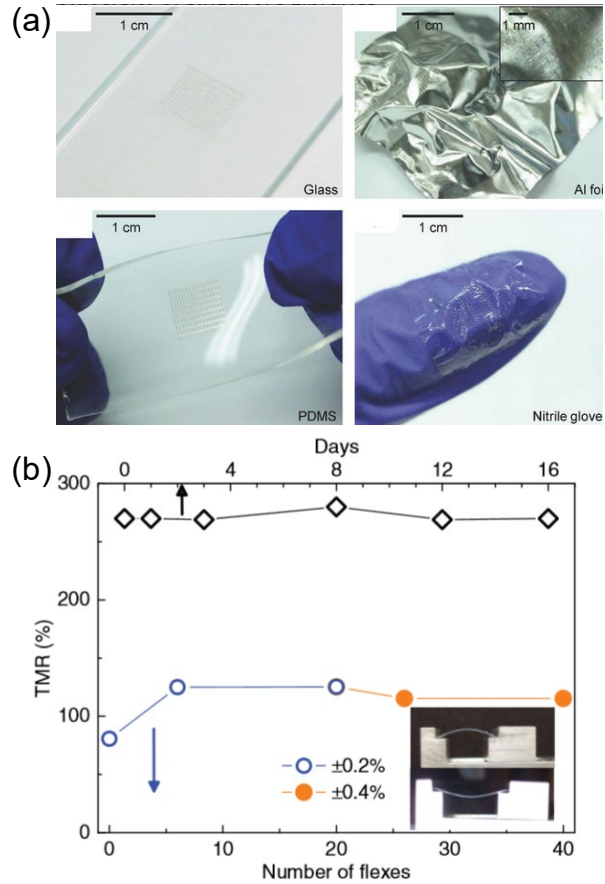

Fig. 26. (a) MTJ transferred on different types of flexible substrates. (b) Robustness of TMR for MTJ on a flexible PET substrate. The bottom axis represents the number of flexes applied using the setup shown in the inset. The top data are for a different MTJ tested over several days (top axis). Reprinted figure with permission from [256]. [copyright statement].

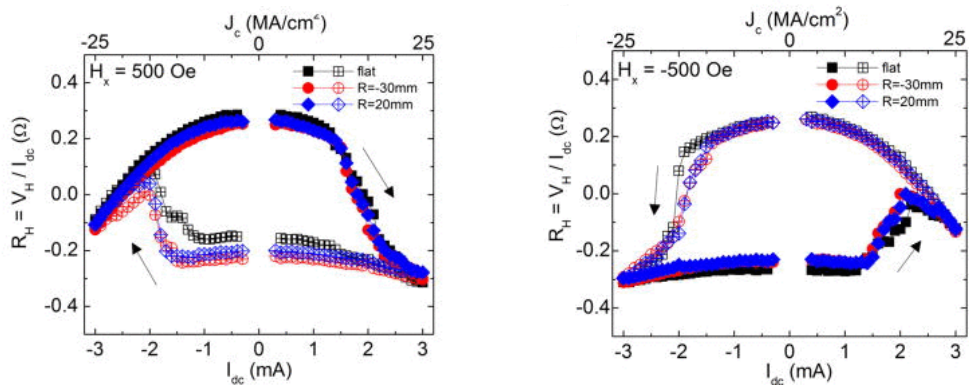

Fig. 27. Robust current-induced switching for SOT device grown on flexible plastic substrate for different bending conditions. Reprinted figure with permission from [263]. [copyright statement].

## VI. Spintronics in terahertz

The electromagnetic (EM) radiations with frequency in the range 100-300 GHz to 3-30 THz are termed as terahertz (THz) radiations, which find applications in spectroscopy, medical imaging, communication etc. A low cost and energy-efficient THz source/emitter is desirable to fully develop these THz systems and further expand their applications. The currently used THz emitters based on photoconductive semiconductors, electro-optic crystals (e.g. ZnTe), air plasma based emitters, etc. have drawbacks in terms of narrow bandwidth, skipped bandwidth or requisite of a high energy laser pump. In view of these shortcomings, there has recently been wide-ranging research interest to explore spintronics based THz emitters.

THz emission using spintronic devices is related to the ultrafast spin dynamics which was at first revealed in sub-picosecond demagnetization in Ni using a femtosecond laser pulse [264]. It has been proposed that a superdiffusive transport of spin-polarized electron is responsible for this ultrafast demagnetization [265]. This mechanism was later confirmed by an experiment involving the Ni and Fe layer with the parallel and antiparallel alignment between them [266]. When the Ni layer was pumped with a femtosecond laser pulse, an increase (decrease) of Fe magnetization was observed when the two layers were parallel (antiparallel) to each other. Soon after, Kampfrath et al. used Fe/(Au or Ru) bilayers to detect the superdiffusive spin current [267]. The schematic of the scheme they used is shown in Fig. 28(a). The laser-induced superdiffusive spin current ($J_s$) on arrival in the metallic Au or Ru layer is converted to a charge current ($J_c = \theta_{SH} J_s \times M/|M|$) due to ISHE. The charge pulse is converted to a EM wave with a frequency in the THz spectrum, governed by the Maxwell's equation. The THz radiation was electro-optically sampled the results of which are shown in Fig. 28(b). It can be seen that a reversal of the magnetization results in a corresponding reversal of the THz signal due to the reversal in direction of $J_s$. It should be noted that the emitted THz were polarized in the x-direction for the sample magnetization (*M*) along the y-direction in Fig. 28(a).

Since both Au and Ru are not the materials with large $\theta_{SH}$, the emitted THz was of limited amplitude. Seifert et al. and Wu et al. have performed extensive studies about different non-magnetic (NM) and FM combinations to realize efficient, broad-band and high-performance spintronic THz emitters [268,269]. Fig. 29(a) compares the amplitude of emitted THz for NMs adjacent to a CoFeB layer. A combination of Pt/CoFeB emits the strongest THz signal. For a W and Ta sample the emitted THz is of opposite sign compared to others due to the opposite $\theta_{SH}$ of these metals. The THz emitter optimized for the NM and FM thicknesses by Wu et al. is shown in Fig. 29(b,c). An increase in THz signal for increasing HM and FM thickness followed by its attenuation is a result of balance between the limited spin diffusion and THz absorption in the two layers [269]. It has also been proposed that the peak THz signal for a particular thickness is possibly due to constructive Fabry-Pérot interference at this thickness [268].

Seifert et al. showed that a Pt/CoFeB/W heterostructure, optimized for individual layer thickness, emits a very strong and broad THz. Fig. 30 compares the time and frequency domain spectra of this spintronic trilayer with other crystal and semiconductor based emitters. Clearly, the spintronic THz emitter outperforms these traditional emitters (showing limited frequency responses due to phonon resonances), especially in terms of its broad frequency coverage. Multilayers of [Pt (2 nm)/Fe (1 nm)/MgO (2 nm)]$_n$ with different repetition numbers also serves as an excellent THz source [270]. A peak THz emission was found for three repetitions of these layers. Apart from conventional HMs, exotic materials such as a TI $Bi_2Se_3$ [271] and monolayer $MoS_2$ [272] in combination with Co have been demonstrated to emit sizeable THz waves (Fig. 31). Attachment of a collimating Si lens was proposed to collect most of the diverging THz to maximize the power output [273]. The THz can also be enhanced by passing currents through the heterostructure resulting in an additional photoconduction related contribution to the total THz [274].

While the initial reports on THz generation assume an essential presence of net magnetization in the system for a finite THz generation, recent THz emitter reports based on nearly-compensated ferrimagnets have proved it otherwise. Chen et al. have demonstrated emission of a finite THz signal from a nearly-compensated $Co_{1-x}Gd_x$/Pt based heterostructure [275]. The magnitude of this THz is comparable to the pure Co based emitter. In fact, the polarity of emitted THz reverses when the composition of $Co_{1-x}Gd_x$ traverses from a Co rich to Gd rich state as shown in Fig. 32(a,b). This behavior of RE-TM base THz emitter is due to the localized nature of magnetic moment carrying f-shell electron in the RE metals. Therefore, the contribution to the superdiffusive spin current is only from the Co sub-lattice. Similar results were reported by Schneider et al. for another RE-TM ferrimagnet, $Fe_{1-x}Tb_x$ [276]. It was found that a CoGd heterostructure emits a stronger THz compared to the CoTb ones possibly due to a large out-of-plane anisotropy in CoTb [277]. The anomalous Hall effect (AHE) as a possible alternate mechanism of THz generation instead of ISHE has been put forward recently [278]. A single FeMnPt layer without a heavy metal was found to generate considerable THz as shown in Fig. 32(c).

To summarize this section, magnetic heterostructures provide a cheap and efficient solution for THz generation. The peak intensity of the generated THz from a NM/FM structure exceeds compared to that from a ZnTe and GaP based emitter (500 μm) which are conventionally used for THz generation. While the ZnTe and GaP THz spectrum shows considerable gaps, specifically between 3 to 13 THz, the spectrum of the spintronic THz emitter is wider and continuous. The THz amplitude for most of the spectral range from spintronic emitters exceeds that of the ZnTe and the GaP based emitters. When compared to a photoconductive switch, the spintronic emitter has a wider bandwidth with a larger intensity above 3 THz. For frequency below 3 THz, a photoconductive switch performs better [268]. A significant enhancement of the THz signals in a lower THz frequency range below 1 THz can be achieved by a novel ultra-broadband spintronic THz emitter enhanced by a current modulation through the semiconductor channel [274].

The spintronics THz emitter films are easy to fabricate and do not require any high temperature deposition process or specific substrate. Future works on spintronic THz emitters should be focused on further improvement of the THz signal at lower laser fluence, removal of the external magnetic field and enabling robustness to external temperature and magnetic fields.

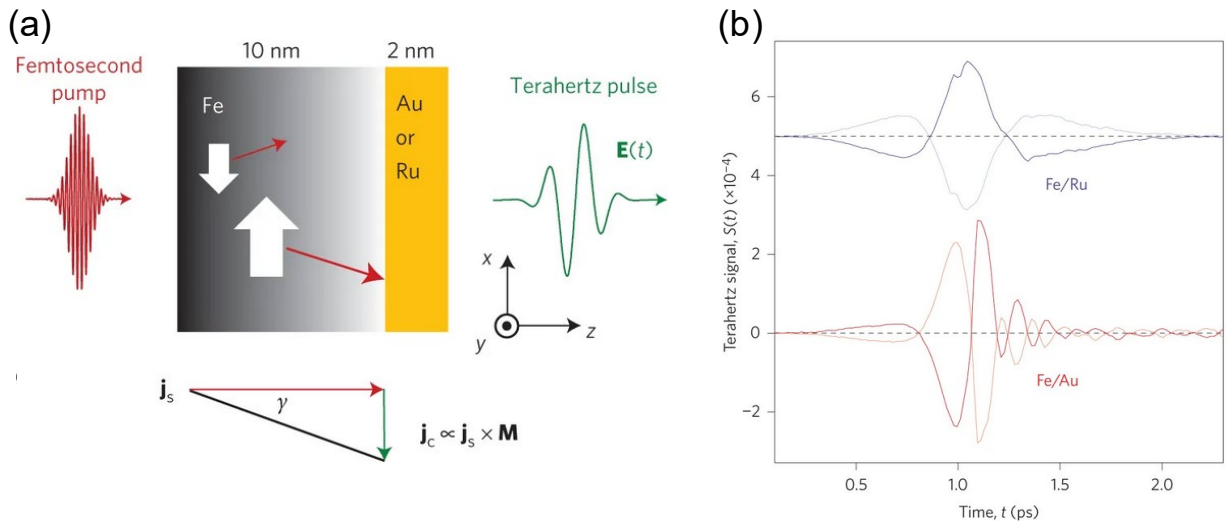

Fig. 28. (a) Mechanism of THz emission using a magnetic/non-magnetic heterostructure and a femtosecond laser pulse. (b) Transient THz response from a Fe/Ru and Fe/Au samples. Reprinted figure with permission from [267]. [copyright statement].

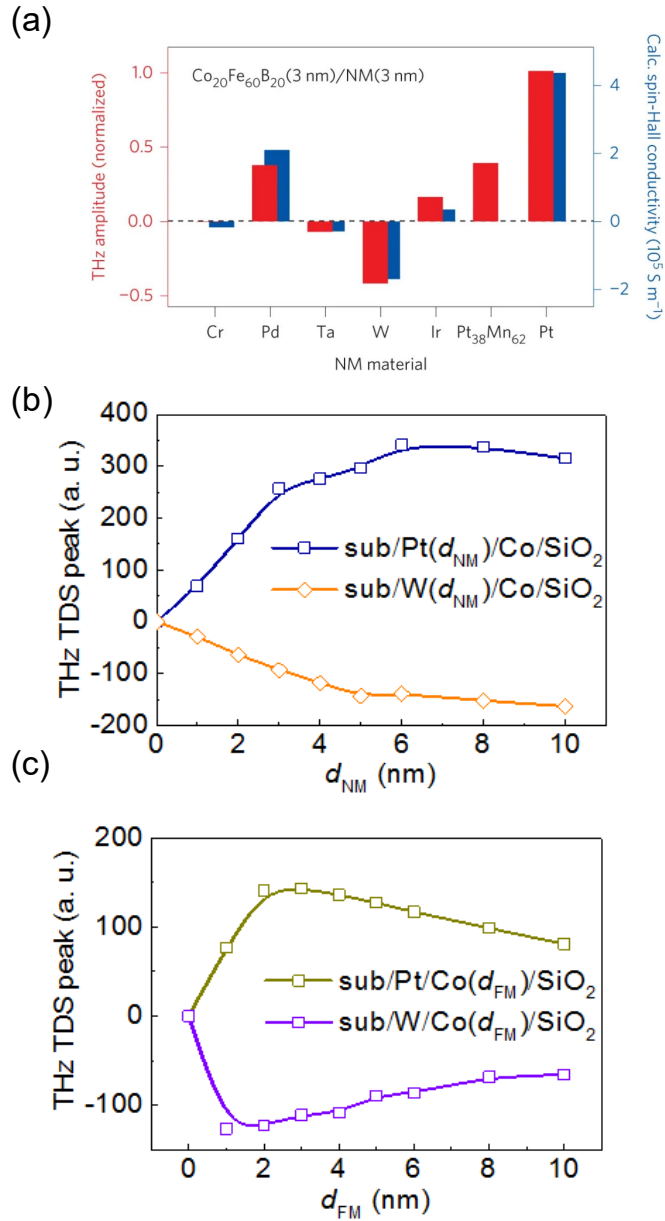

Fig. 29. (a) Emitted THz amplitude for samples with combination of different non-magnets (NM) with CoFeB (red bar). The blue bar represents the calculated spin Hall conductivity for these materials. Reprinted figure with permission from [268]. [copyright statement]. Amplitude of emitted THz as a function of (b) NM and (c) Co thickness for NM/Co samples. Reprinted figure with permission from [269]. [copyright statement].

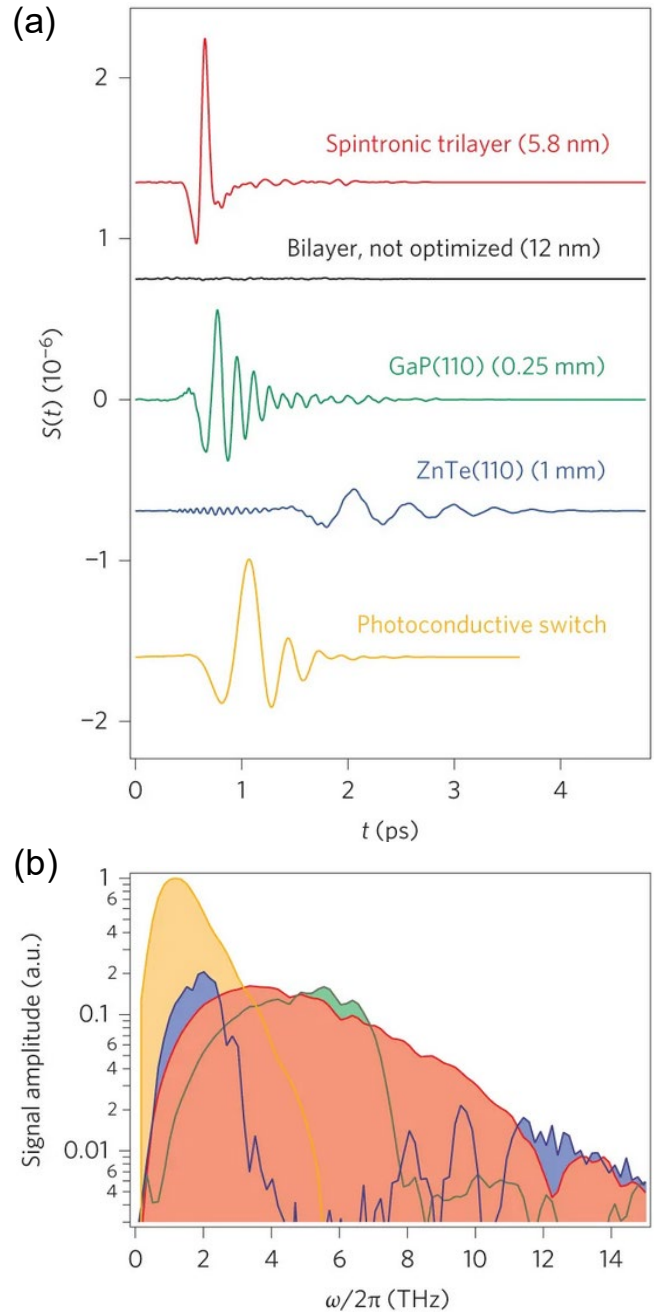

Fig. 30. THz signal in the (a) time and (b) frequency domain for different THz emitters. The color legend in (b) is same as shown in (a). Reprinted figure with permission from [268]. [copyright statement].

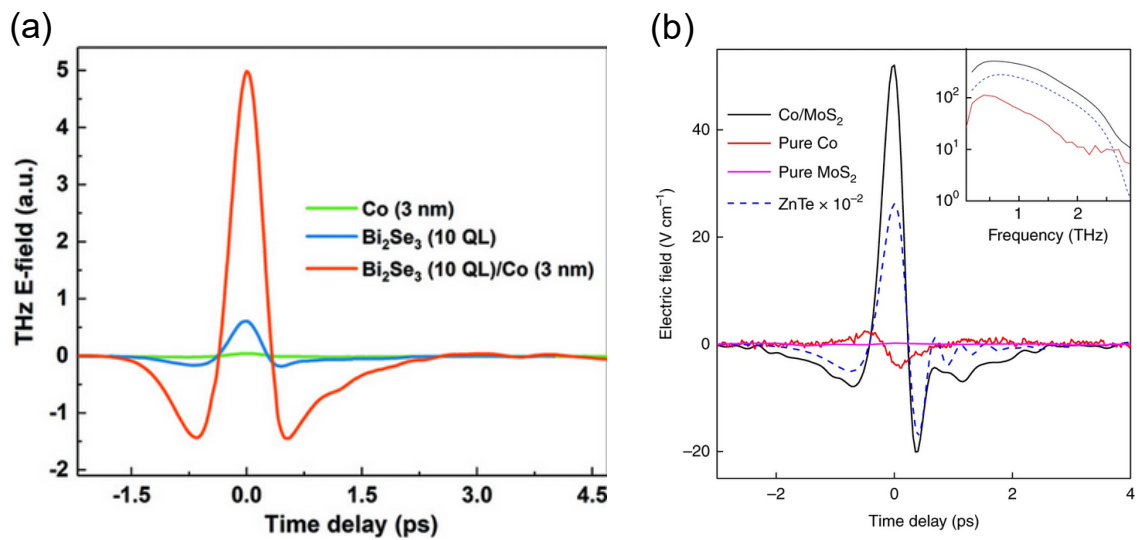

Fig. 31. THz signal from a (a) topological insulator (Reprinted figure with permission from [271]. [copyright statement]) and (b) monolayer $MoS_2$. Reprinted figure with permission from [272]. [copyright statement].

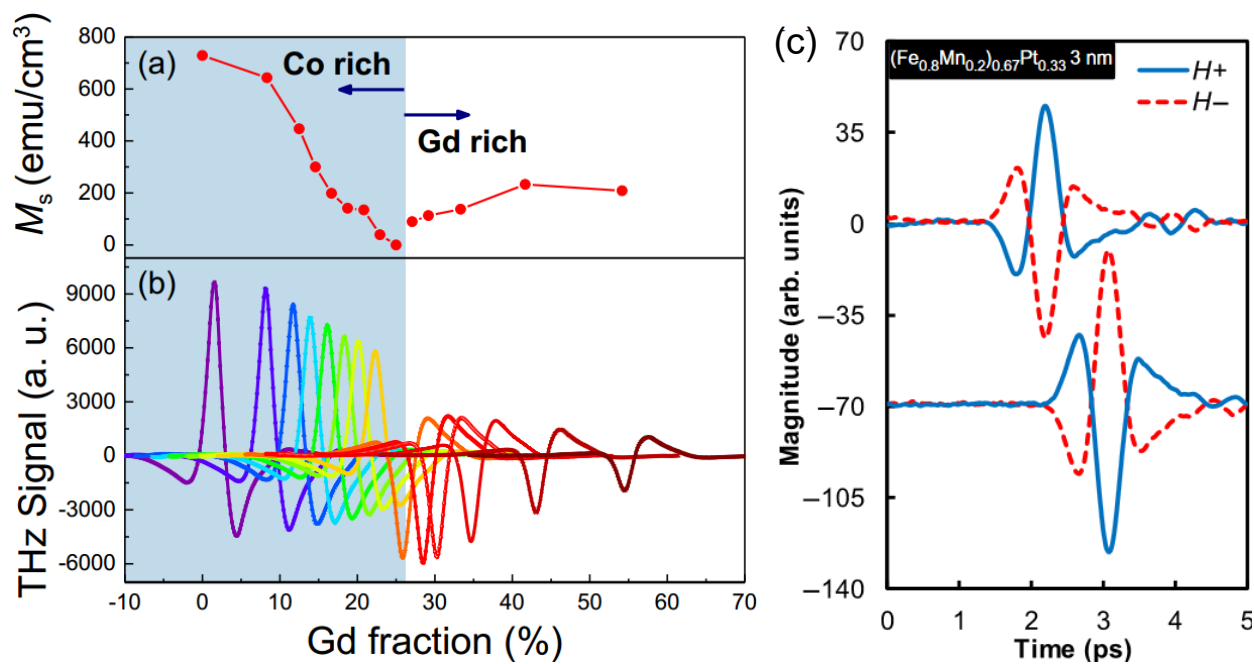

Fig. 32. (a) Magnetization of the samples as a function of Gd composition. (b) The emitted THz for the samples. Reprinted figure with permission from [275]. [copyright statement]. (c) Transient THz response from a single ferromagnetic layer. The waveform in the top panel is obtained from shinning the femtosecond laser pulse on the quartz side (substrate) and the bottom waveform is on shinning it from the MgO (capping) side. Reprinted figure with permission from [278]. [copyright statement].

## VI. Outlook

Ever since its use in the magnetic-core memory, spin devices have come a long way to be used in nanometer dimensioned STT memories. Apart from forming backbone of mass storage in form of hard disk, almost all the modern-day devices and instruments are packed with magnetic sensors of various types. With the end of Moore's law in sight, finding replacement of silicon logic devices is at the forefront of solid-state device research. While there is a continuous stride towards development of spin-based logic devices as discussed in this review, the spin logic research is still in a very preliminary stage. Currently most of these spin logic devices have been demonstrated in form of stand-alone gates or logic elements. The viability of a complete spin logic architecture can only be ascertained if the individual spin elements are suitably integrated. The future direction of spin logic research should aim towards this assimilation. Apart from this, for all practical purposes any logic device should not only perform the said logic function, but it should be capable of meeting other important parameters such as low power, large fan-in, large fan-out, high speed, etc. Future efforts on spin logic devices should focus on developing devices that meet all of these criteria. Overall, spin logic circuits require continuous research effort to become practical in the distant future.

Spintronic memories or MRAM have proven to be one of the most successful spintronic applications. In fact, they are one of the most promising among all the non-volatile memory candidates currently being pursued. The second generation MRAM i.e. STT-MRAM is already mass produced and serving customer needs in companies such as Everspin, GlobalFoundries, and Samsung among those involved in its active production and development. The next iteration of MRAM involving the use of SOTs is a focus of major research in both academia and industry due to their promising high speed and large endurance. The MRAMs are currently developed as standalone memories and for embedded applications. One of the target applications of MRAM is in a form of cache to enable low power computing. However, STT-MRAMs at their few to tens of ns read speed still fall short of the speed requirements of the L1/L2 cache. The promising high speed of SOT-MRAM due to low incubation times makes them a competitive candidate for L1/L2 cache. However, currently one hindrance towards practical applications of SOT-MRAM is the requirement of external magnetic field for their deterministic switching operation. While many device structures and engineering solutions have been proposed to overcome this problem, a full-scale integration of these SOT device with CMOS compatible process is yet to be seen. The future research work on MRAM should also focus on device scaling and development of a high density MRAM architecture. MRAMs outperform both the conventional non-volatile solid-state storage like eFLASH and standalones memories i.e. DRAM in all performance aspects other than their storage density. This makes MRAMs cost ineffective as of now. Future works towards improving their storage density will definitely make them a lot more competitive in a broader range of the memory pyramid. In addition, the potential of other magnetic memory candidates such as skyrmions should also be continuously evaluated.

The field of solid-state devices for non-von Neumann computing itself is in a very early stage with majority of success being shared by memristors. However, recently there have been continuous demonstrations of non-von Neumann systems using spin devices. This suggests the viability of spintronics as one of the promising approaches for pursing alternative computing methodologies [231,233,234]. Being a field under development, alternative computing schemes do not yet have a coherently defined requirements from the devices and systems. In both memristor and spintronics, the researchers, with the tools in their hands, are proposing various standalone devices and architectures which although solving the specific problem in question, fall short of marching towards a coherent and a general-purpose hardware for alternative computing needs. Since the viability of spintronic devices for ANNs, biological neural network and other computing schemes has been already suggested, the next step should be towards a more joint effort between the device, circuits and system architecture teams for obtaining tangible spin solutions. At present, the ANNs and ANN-like systems made from memristors and spintronics perform only the inference step, while the learning and sometimes even weight storage is still carried out in conventional computers. Going forward, while it is desired to move the learning tasks to the new solid-state devices, a hybrid architecture can be a more practical option. Like spin-logic, the real application of spins in unconventional computing will become clearer in the coming decade.

As detailed in this review, spintronics should not only be approached as a device candidate for computing applications, but its physics should also be exploited for non-computing systems. In form of a THz emitter, the spin devices have already proved themselves very competitive. A fast track development of THz systems based on these devices can be in fact applied to some applications in the coming few years. While in this review we have discussed some of the major emerging applications of spintronics, there are analogous ongoing research efforts for several other equally important applications e.g. spin based electronic oscillators [279-281] which may find their use in both communication and computing. The low complexity and less stringent device requirement of non-computing systems should aid spintronics researches to quickly transfer a research device from an academic laboratory to industry, which requires not only a functionality testing of an individual device, but also manufacturability and scalability aspects. Finally, flexible electronics will be playing a big role in the future, especially in consumer electronics. Works of any spin device on rigid wafers such as silicon for both computing and non-computing needs should have an analogous effort towards developing them on flexible substrates.

**VII. Conclusion**

In this review, we have discussed the novel emerging areas of spintronic applications. Spin devices proposed for the logic computation were discussed in Section II. The low static power consumption of spin elements makes them attractive for logical devices. However, full replication of the speed and versatility offered by the CMOS is one of the foremost challenges in this area of research. In Section III, spintronic memories were discussed with emphasis on the latest generation of spin memories i.e. SOT-MRAM. The SOT which requires a simpler device design promises larger endurance and smaller energy consumption when compared to the currently commercialized STT-MRAM. Current efforts to improve the energy-efficiency of SOT devices involve engineering the SOC source, magnetic layer and the heterostructure stack design itself. In addition, oxygen incorporation in various layers has proved to an effective way to modulate and enhance the SOTs. Alternative computing methodologies not based on Von Neumann architecture are currently being pursued actively for their critical future importance. In Section IV, we discussed the recent research progress in this area with respect to the emerging spin devices that mimic the biological synapse and neuron functions. The optimization and recognition systems based on MTJs have been proposed to be both area and energy effective when compared to similar implementation using CMOS. The viability of spintronics devices for flexible electronics was discussed in Section V. Several spintronic devices such as GMR, TMR, SOT, exchange devices, etc. have been shown to perform robustly on variety of flexible substrates such as plastic, Kapton tape, PDMS, PET, etc. In the final section (Section VI), we detailed on the advancement of cost effective THz emitter based on magnetic/non-magnetic heterostructures. The THz emitted from these spin devices are of equivalent strength and much broader bandwidth compared to the conventional crystal and semiconductor-based emitters.

Overall we see that spintronics has emerged as a very promising and an actively pursued solid-state technology for meeting the future (opto)electronics needs in a wide variety of application areas. Future spintronic research should focus on improvement of spin device on all fronts that concern the stringent requirement of device practicality. In addition, research concerned towards system level implementation of spintronics should also be pursued.

**ACKNOWLEDGMENT**

H.Y. is grateful to the IEEE Magnetics Society, supporting as a Distinguished Lecturer in 2019 to discuss major portion of this work with many research groups worldwide. The work is partially supported by SpOT-LITE program (A*STAR grant, A18A6b0057) through RIE2020 funds, Singapore Ministry of Education (MOE) Tier 1 (R263-000-D61-400-114), Samsung Electronics' University R&D program (Exotic SOT materials/SOT characterization), AME-IRG (Grant No. A1983c0037) through RIE2020 funds, and NUS Hybrid-Integrated Flexible (Stretchable) Electronic Systems Program.